# Curved Monopoles and Yagi–Uda Arrays for Low-Angle HF Skywave Radiation over Lossy Ground: Analysis, Hybrid Bayesian Optimization, and Validated Full-Wave Design

**MASOUD SALMANI ARANI**[1], (Graduate Student Member, IEEE), **REZA SHAHIDI**[1], (Senior Member, IEEE), **and LIHONG ZHANG**[1], (Senior Member, IEEE)

[1]Department of Electrical and Computer Engineering, Memorial University of Newfoundland, St. John's, NL A1C 5S7, Canada

Corresponding author: Masoud Salmani Arani (e-mail: msalmaniaran@mun.ca).

This work was supported in part by Natural Sciences and Engineering Research Council of Canada (NSERC).

**ABSTRACT** High-frequency (HF) skywave radars require high gain at take-off angles of 3°–30°, where vertically polarized monopoles and Yagi–Uda arrays over lossy land suffer from pseudo-Brewster cancellation. This paper shows that curving the elements of HF monopoles and Yagi–Uda arrays converts radiation from this soil-limited transverse-magnetic route into a transverse-electric route that is nearly insensitive to soil conditions. Closed-form half-space bounds quantify both routes and yield design rules; a separability result extends the soil insensitivity to arrays. A hybrid design method combines Bayesian optimization on analytical and method-of-moments models with full-wave simulations of a screened site on moist soil; current extraction differs by at most 5.1% RMS, and Fresnel post-processing agrees with independent perfect-ground results within 0.16 dB in pointwise gain for four arrays. At 15 MHz, a self-matched 2-element curved Yagi–Uda array and a Bayesian-optimized 3-element array achieve realized forward-sector gains of 6.8 and 8.3 dBi over moist soil, 6.4 and 6.3 dB above straight quarter-wave Yagi–Uda arrays optimized by the same procedure, with 92%–95% efficiency. Their gain changes by about 1 dB from a perfect ground to very dry soil, whereas that of the straight arrays falls by 9 dB. Bayesian optimization, confirmed by full-wave simulation, identified a design 0.2 dB above the sweep optimum and a gain–bandwidth Pareto front that offers up to twice the matched bandwidth. The results also show that the native far field of full-wave solvers can overestimate the low-angle gain of vertical arrays over soil by more than 4 dB.

## I. INTRODUCTION

High-frequency (HF) radars operating at 3–30 MHz achieve over-the-horizon (OTH) coverage either by skywave propagation, which uses the ionosphere to reach ranges of thousands of kilometers, or by surface-wave propagation along the sea, which is limited to a few hundred kilometers [1]–[4]. Skywave systems illuminate the ionosphere at take-off angles between about 3° and 30°, e.g., 14°–21° for a 2090-km path [5]. Because the link budget of a skywave radar scales with the product of the transmit and receive gains in this sector, every decibel gained at these angles translates directly into effective radiated power and two-way signal-to-noise ratio.

The long wavelength makes HF antenna design a trade-off between gain, bandwidth, and size. Electrically-small and top-loaded monopoles suffer from high reactance and low efficiency [6], [7], whereas log-periodic, dipole, and cavity-backed designs need tall supports or added complexity [8]–[10]. Vertically polarized monopoles over a ground screen and the vertical Yagi–Uda arrays built from them remain the most common radiators, but over land their low-angle gain is limited by the ground itself: below the pseudo-Brewster angle the reflected wave cancels the direct wave [11]–[13]. Radial or mesh screens reduce the near-field loss [14], but they cannot alter the far-field reflection, which forms hundreds of meters from the antenna. Coastal siting is therefore preferred [2], yet many transmit sites lie inland, where the soil moisture, and with it the ground constants, change with season.

In previous work, we introduced a partially-curved, approximately-quarter wave monopole for HF radar,

comprised of a straight vertical tube supporting an arc, and showed on a perfect electric conductor (PEC) plate that curvature improved the match, widened the bandwidth, and increased the peak realized gain [15]. That study left open whether the benefit survives over lossy soil, which governs the low-angle reflection at real sites, what physical mechanism is responsible, how curvature should scale with the element length, and whether curved elements can be combined into directive arrays. Three further requirements guide the present work: the predicted gain must not rely on an assumed current distribution, but on currents validated independently over lossy ground; the radiator should be directive, as required by skywave radars; and the comparison must include straight antennas optimized for each ground by the same procedure. This paper meets these requirements and extends the curved element to curved Yagi–Uda arrays.

Wire antennas over lossy ground are classically analyzed using the reflection-coefficient (Fresnel) approximation of the Sommerfeld problem [11], [13], [16], which is accurate in the space-wave region when the Norton surface wave is excluded, and method-of-moments (MoM) codes such as NEC [17]. Full-wave finite-element solvers model the near-field interaction with the soil and the ground screen in detail, but their far-field transformation assumes free space beyond the solution box and therefore cannot reproduce the ground reflection that forms hundreds of wavelengths away. Yagi–Uda arrays [18] are designed with well-established rules for straight elements [19], but, to the authors' knowledge, the design of Yagi–Uda arrays of curved, ground-mounted elements, and the interplay between their polarization and the soil, have not been reported. Bayesian optimization (BO) with Gaussian-process surrogates [20]–[22] is well suited to expensive electromagnetic objectives, and trust-region variants [23] handle higher-dimensional problems efficiently, but their effectiveness depends on an objective that is both fast and physically correct.

The main contributions of this paper are as follows:

1) **Analysis:** Closed-form bounds derived from the space-wave field over a half-space separate a soil-limited transverse-magnetic (TM) route from a nearly soil-insensitive transverse-electric (TE) route (Lemmas 1–2, Propositions 1–3). A separability result (Proposition 4) shows that this insensitivity carries over to arrays of curved elements that share a common mast and arc circle, which motivates a new construction rule for curved Yagi–Uda arrays.

2) **Methodology:** A hybrid design framework combines BO on fast analytical and thin-wire MoM models with full-wave HFSS simulations of a screened site on lossy soil. The framework includes a current-extraction procedure with an RMS difference no greater than 5.1%, a closed power balance for the near-field efficiency, and Fresnel post-processing validated against exact perfect-ground solutions within 0.16 dB in pointwise gain for four arrays. The framework also quantifies how far the native far field of the full-wave solver departs from the physically correct low-angle gain over soil.

3) **Novel radiators:** A self-matched 2-element and a Bayesian-optimized 3-element curved Yagi–Uda array achieve realized forward-sector gains of 6.8 and 8.3 dBi over moist soil, 6.4 and 6.3 dB above straight quarter-wave Yagi–Uda arrays optimized by the same procedure, while their gain varies by only about 1 dB from a perfect ground to very dry soil.

4) **Fair and reproducible comparison:** All baselines are optimized with the same budget and selection rule, the effect of curvature alone is isolated with straight arrays of identical element lengths, and robustness to ±2% construction errors is quantified. BO is shown to deliver an HFSS-confirmed Pareto front between gain and bandwidth, whose high-gain end outperforms the best parametric-sweep design.

The overall analysis, design, optimization, and validation workflow is summarized in Fig. 1. The rest of this paper is organized as follows. Section II presents the analysis. Section III describes the full-wave site model, the post-processing, and the optimization framework. Section IV summarizes results from the single curved element. Section V presents the design and optimization of the curved Yagi–Uda arrays, Section VI their validation, and Section VII the comparison with straight arrays over seven grounds. Section VIII discusses robustness, practical implications, limitations, and the planned scaled measurements, and Section IX concludes the paper.

## II. ANALYSIS

### A. SPACE-WAVE FIELD OF A CURVED MONOPOLE

Consider a thin wire in the $xz$ plane, as depicted in Fig. 2(a). A vertical tube of height $h_t$, fed against a ground screen, supports an arc of radius $R$ and angle $\theta_c$ that bends toward the $-x$ direction; the conductor length is $L = h_t + R\theta_c$. Let $\ell$ be the arc length, $(x, 0, z)$ the corresponding position on the wire, $(t_x, 0, t_z)$ the unit tangent, and $I(\ell)$ the current. The soil is a half-space of complex relative permittivity $\varepsilon = \varepsilon_r - j\sigma/(\omega\varepsilon_0)$. For elevation $\psi$ and azimuth $\phi$ ($\phi = 0$ along $+x$), let $s = \sin\psi$, $c = \cos\psi$, and $q = (\varepsilon - c^2)^{1/2}$ with $\mathrm{Re}\, q > 0$. Excluding the Norton surface wave, which decays faster than $1/r$ [11], [12], the far field is

$$E_\theta = \kappa \int_0^L I(\ell)\, e^{jkxc\cos\phi}\big(s\, t_x \cos\phi\, \tilde{T}_v - c\, t_z T_v\big) d\ell \tag{1}$$

$$E_\phi = -\kappa \sin\phi \int_0^L I(\ell)\, e^{jkxc\cos\phi}\, t_x T_h\, d\ell \tag{2}$$

where $\kappa = -jk\eta_0 e^{-jkr}/(4\pi r)$, $T_v = e^{jkzs} + \Gamma_v e^{-jkzs}$, $\tilde{T}_v = e^{jkzs} - \Gamma_v e^{-jkzs}$, $T_h = e^{jkzs} + \Gamma_h e^{-jkzs}$, and $\Gamma_v$ and $\Gamma_h$ are the Fresnel coefficients [13], [24]. Expressed through $q$, they satisfy the exact identities

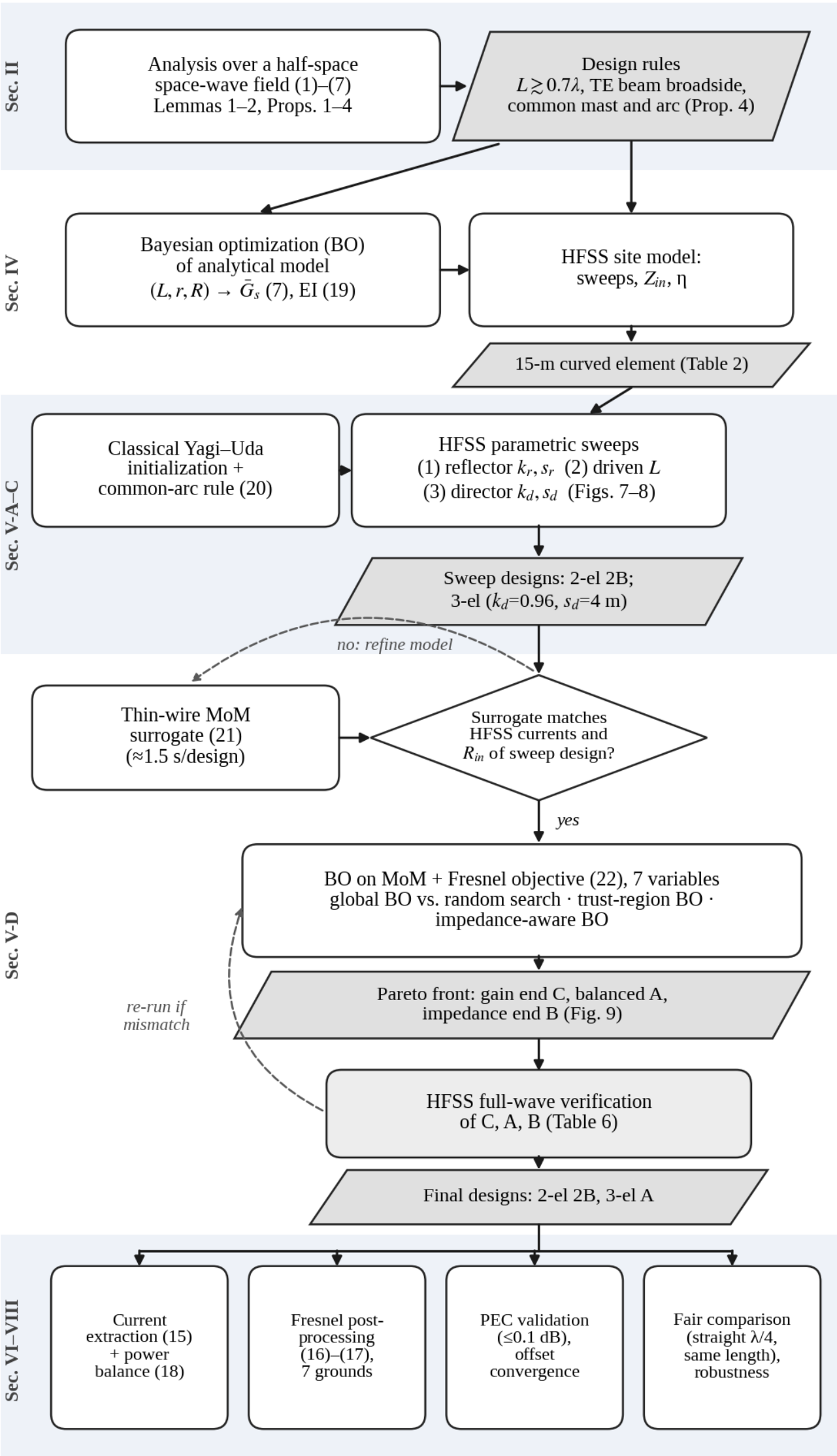

**FIGURE 1.** Design, optimization, and validation flow of this paper, ordered by section: closed-form analysis (Section II); BO of the analytical single-element model followed by full-wave refinement (Section IV); knowledge-based initialization and HFSS parametric sweeps of the arrays (Sections V-A–V-C); BO on a validated MoM surrogate with full-wave verification of the Pareto candidates (Section V-D); and current-based evaluation, validation, and comparison (Sections VI–VIII). Parallelograms denote results, the diamond a decision, and dashed arrows feedback loops.

$$1+\Gamma_v = \frac{2\varepsilon s}{\varepsilon s + q}, \qquad 1-\Gamma_v = \frac{2q}{\varepsilon s + q} \tag{3}$$

$$1+\Gamma_h = \frac{2s}{s + q}, \qquad 1-\Gamma_h = \frac{2q}{s + q} \tag{4}$$

which yield

$$T_v = 2\cos(kzs) - (1-\Gamma_v)e^{-jkzs} \tag{5}$$

$$T_h = 2j\sin(kzs) + (1+\Gamma_h)e^{-jkzs} \tag{6}$$

The first terms of (5) and (6) are the perfect-ground (PEC) values, and the second terms are the exact soil corrections. The figure of merit is the solid-angle mean gain over the skywave sector $\Omega_s = \{3^\circ \le \psi \le 30^\circ\}$,

$$\bar{G}_s = \frac{1}{|\Omega_s|}\int_{\Omega_s} G\, d\Omega = \bar{G}_s^{TM} + \bar{G}_s^{TE} \tag{7}$$

where the split into the $E_\theta$ (TM) and $E_\phi$ (TE) contributions is exact because the two components are orthogonal. For a single element, $\Omega_s$ covers all azimuths; for directive arrays, we also use the forward sector $\Omega_f = \{3^\circ \le \psi \le 30^\circ,\ 45^\circ \le \phi \le 135^\circ\}$ and denote the corresponding mean gain by $\bar{G}_f$.

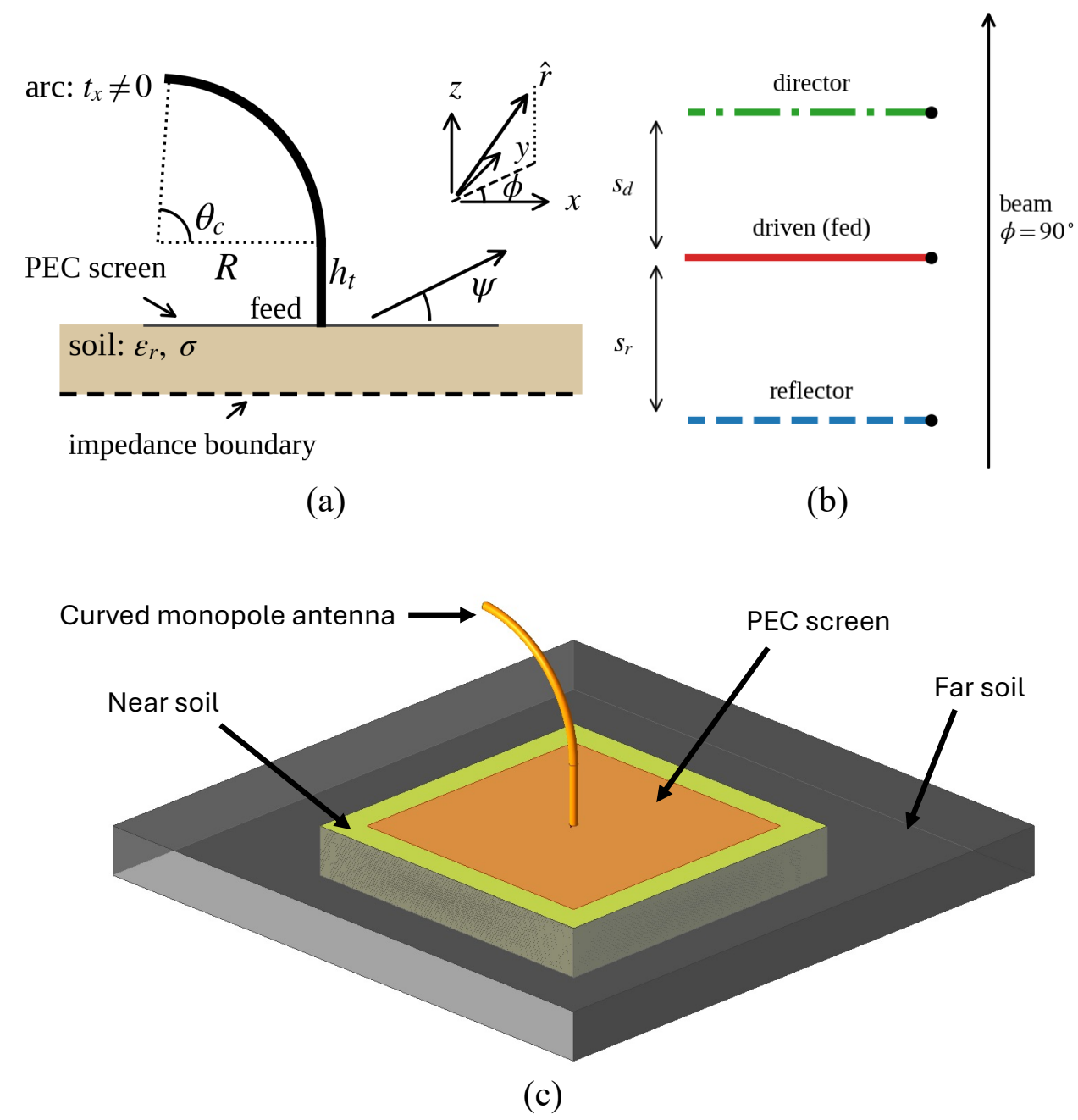

**FIGURE 2.** (a) Curved monopole on a PEC ground screen over a lossy-soil half-space (side view). Elevation $\psi$; azimuth $\phi$ from $+x$; the arc bends toward $-x$. (b) Top view of the curved Yagi–Uda array: reflector, driven element, and director lie in parallel planes at $y = -s_r$, 0, and $s_d$; the beam points toward $+y$ ($\phi = 90^\circ$). (c) Perspective view of the HFSS site model: element, PEC screen (top square surface), near-soil region (inner block beneath the screen), and far-soil region (outer block).

### *B. BOUNDS ON THE TWO POLARIZATION ROUTES*

*Lemma 1 (TM ground factor):* Let $|\varepsilon| > 1$, $s_B = (|\varepsilon| - 1)^{1/2}/|\varepsilon|$, and $s = \nu s_B$ with $0 < \nu < 1$. Then

$$\frac{|\varepsilon| s}{|\varepsilon| s + \sqrt{|\varepsilon| + 1}} \le \frac{|1+\Gamma_v|}{2} \le \frac{\nu}{1-\nu} \tag{8}$$

*Proof:* Since $|\varepsilon - c^2| \ge |\varepsilon| - 1$, we have $|q| \ge (|\varepsilon| - 1)^{1/2} = |\varepsilon| s_B$, so that $|\varepsilon| s = \nu |\varepsilon| s_B \le \nu |q|$ and $|\varepsilon s + q| \ge |q| - |\varepsilon| s \ge (1-\nu)|q|$. The upper bound then follows from (3). For the lower bound, $|\varepsilon s + q| \le |\varepsilon| s + |q|$ and $|q|^2 = |\varepsilon - c^2| \le |\varepsilon| + 1$. ∎

Hence the TM ground factor rises from zero to order unity across $s \approx s_B \approx |\varepsilon|^{-1/2}$. For the soils of Table 1, $\arcsin s_B$ matches the exact pseudo-Brewster angle $\psi_B$ within 0.6° (1.9° for very dry soil). Fig. 3(a) plots $|1+\Gamma_v|/2$ with the upper bound of (8).

*Lemma 2 (TE ground factor):* For $\sqrt{|\varepsilon| - 1} > s$,

$$|T_h - 2j\sin(kzs)| = |1+\Gamma_h| \le \frac{2s}{\sqrt{|\varepsilon|-1}-s} \qquad (9)$$

*Proof:* The equality follows from (6). By (4), $|1+\Gamma_h| = 2s/|s+q|$, and $|s+q| \ge |q| - s \ge \sqrt{|\varepsilon|-1} - s$. ∎

The two lemmas expose an asymmetry that underlies the entire paper. By (5), the TM correction satisfies $|1-\Gamma_v| \to 2$ at grazing incidence and cancels the PEC term, whereas by (6) and (9) the TE correction vanishes there and is of order $|\varepsilon|^{-1/2}$ elsewhere (Fig. 3(b)).

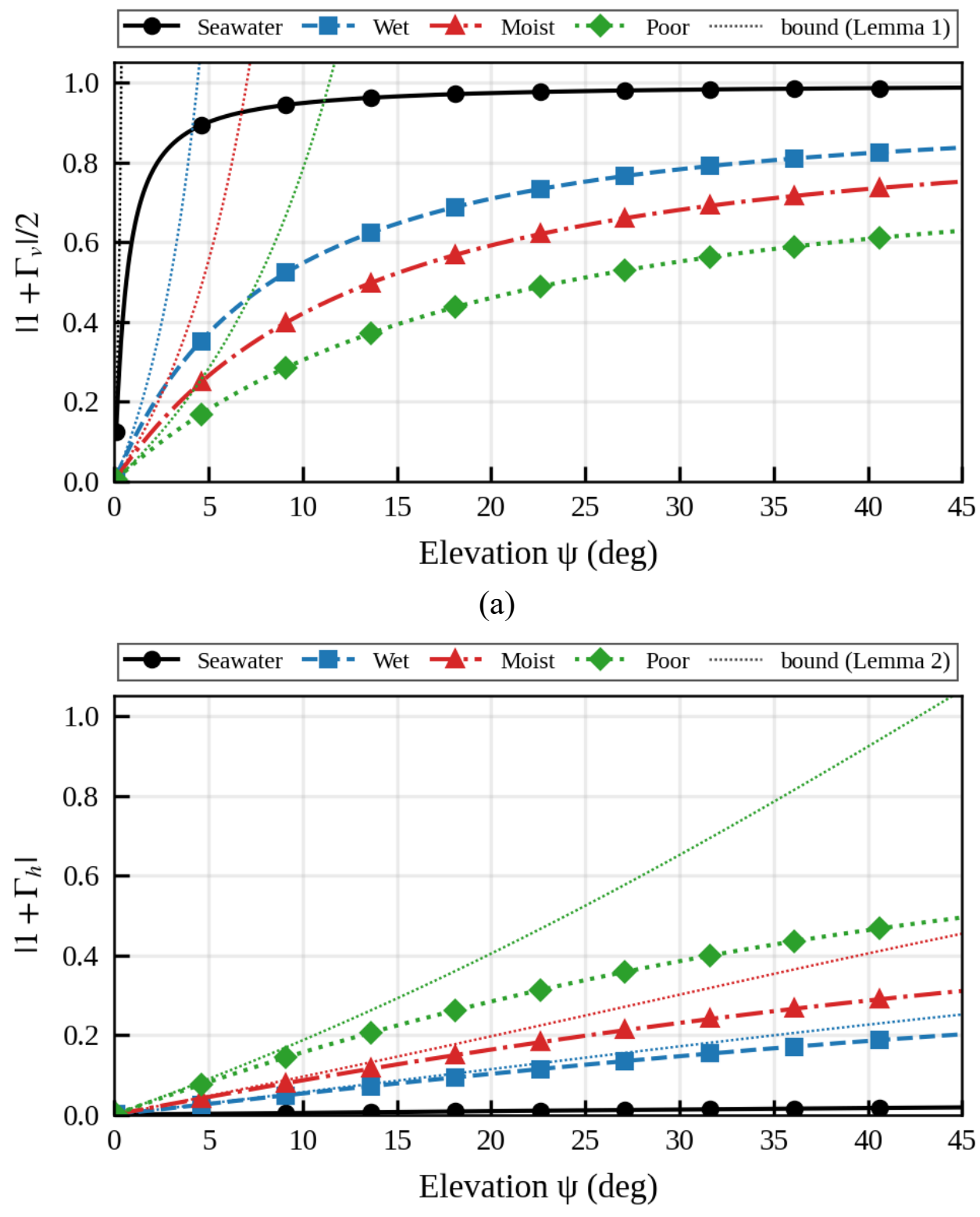

**FIGURE 3.** Ground factors versus elevation for four grounds: (a) TM factor $|1+\Gamma_v|/2$ with the upper bound of Lemma 1 (dotted) and $\arcsin s_B$ (dashed); (b) TE correction $|1+\Gamma_h|$ with the bound of Lemma 2 (dotted).

*Proposition 1 (TM suppression):* For a current element at height $z$ and $s = \nu s_B$,

$$|T_v| \le \frac{2(\nu + kzs)}{1-\nu} \qquad (10)$$

*Proof:* Write $T_v = (1+\Gamma_v)\cos(kzs) + j(1-\Gamma_v)\sin(kzs)$. By Lemma 1, $|1+\Gamma_v| \le 2\nu/(1-\nu)$; the proof of Lemma 1 also gives $|1-\Gamma_v| = 2|q|/|\varepsilon s + q| \le 2/(1-\nu)$; and $|\sin(kzs)| \le kzs$. ∎

Relative to its PEC value of 2, the TM factor of any low element is thus suppressed below $s_B$ regardless of geometry. For moist soil ($s_B = 0.245$) and a 4.67-m quarter-wave element, (10) guarantees at least 8.6 dB of suppression at $\psi = 3°$.

*Proposition 2 (reduced TE sensitivity):* At broadside ($\phi = \pi/2$), let the horizontal current $I(\ell)t_x(\ell)$ have uniform phase and one sign on the arc, which occupies heights $z \in [z_1, z_2]$ with $kz_2 s < \pi$, and let

$$\delta = \frac{s}{\left(\sqrt{|\varepsilon|-1}-s\right)\min_z \sin(kzs)} \qquad (11)$$

Then

$$1-\delta \le \frac{|E_\phi|}{|E_\phi^{PEC}|} \le 1+\delta \qquad (12)$$

*Proof:* By (2) and (6), $E_\phi - E_\phi^{\text{PEC}} = -\kappa \int I t_x (1+\Gamma_h) e^{-jkzs} d\ell$, whose modulus is at most $|\kappa||1+\Gamma_h| \int |I t_x| d\ell$. With uniform phase and one sign, $|E_\phi^{\text{PEC}}| = 2|\kappa| \int |I t_x| \sin(kzs) d\ell \ge 2|\kappa| \min_z \sin(kzs) \int |I t_x| d\ell$. Apply (9) and the triangle inequality. ∎

*Proposition 3 (structure):* (i) By (2), $E_\phi \equiv 0$ for $\phi = 0, \pi$; the TE radiation of any planar curved monopole vanishes in the plane of the wire and peaks broadside. (ii) Over PEC, the horizontal-current terms in (1) and (2) are $O(s^2)$ and $O(s)$ as $\psi \to 0$, whereas the vertical term tends to $-2\kappa \int I t_z e^{jkx\cos\phi} d\ell$; the grazing field is set by the vertical current alone [25]. (iii) For an approximately quarter-wave element with a sinusoidal current, the arc carries the open-end current $|I| \le I_0 \sin(ku)/\sin(kL)$, where $u$ is the distance to the tip, so that $\int_{\text{arc}} |I| d\ell = O(\ell_a^2)$ for arc length $\ell_a$; with $|\sin(kzs)| \le kLs$, the TE power is of fourth order in $k\ell_a$.

*Proof:* (i) follows from the factor $\sin\phi$ in (2). (ii) follows from (1), (2), (5), and (6) with $\Gamma_v = 1$ and $\Gamma_h = -1$, since $\tilde{T}_v = 2j\sin(kzs) = O(s)$ multiplies $s$, and $T_h = 2j\sin(kzs) = O(s)$. (iii) follows from $\sin(ku) \le ku$ and integration over $u \in [0, \ell_a]$. ∎

### C. EXTENSION TO CURVED YAGI–UDA ARRAYS

Consider $N$ planar elements in the planes $y = y_n$, each with its own conductor length and current $I_n(\ell)$. Because (1) and (2) are linear in the current, the array field is the sum of the element fields with the phase factors $e^{jkcy_n \sin\phi}$. The following result explains why the soil insensitivity of the TE route survives in an array.

*Proposition 4 (separability):* Let all elements share the same tube height $h_t$ and the same arc circle (center $(-R, y_n, h_t)$, radius $R$), so that they differ only in their arc angles $\theta_n$, and let their arc currents satisfy $I_n(\ell) = a_n I_1(\ell)$ on the common part of the arcs. Then, on that common part, the broadside TE field factorizes as

$$E_\phi^{array} = E_\phi^{(1)}(\psi; \varepsilon) \sum_{n=1}^{N} a_n\, e^{jkcy_n \sin\phi} \qquad (13)$$

so that the ground enters only through the element factor $E_\phi^{(1)}$, which obeys Proposition 2, while the array factor, and hence the front-to-back ratio (F/B), is independent of the soil.

*Proof:* On the common arc, all elements share $x(\ell)$, $z(\ell)$, and $t_x(\ell)$; hence the integrands of (2) differ only by $a_n$ and the factor $e^{jkcy_n \sin\phi}$, which do not depend on $\ell$. ∎

The assumption $I_n \propto I_1$ holds approximately because the parasitic elements differ from the driven element by only a few percent in length (Section V), and the extracted HFSS currents confirm nearly constant relative phases along the arcs. Proposition 4 motivates the *common-mast, common-arc* construction rule adopted in Section V: it maximizes the overlap of the arcs and therefore preserves the ground insensitivity of the single element. Straight vertical Yagi–Uda arrays obey the analogous factorization for the TM route, but their element factor is the one suppressed by Proposition 1.

### D. CONSEQUENCES AND DESIGN RULES

Curvature converts TM current, whose low-angle field is suppressed below $\arcsin s_B$ (Proposition 1), into elevated TE current, whose field stays within $1 \pm \delta$ of its PEC value (Proposition 2) and peaks broadside at

$$sin\psi^* = \frac{\lambda}{4\bar{z}} \tag{14}$$

where $\bar{z}$ is the current-weighted height of the arc. Four design rules follow. ***Rule 1:*** a soil-mediated advantage over the best straight monopole requires an arc high enough for $\psi^*$ to fall near the window ($\bar{z} \gtrsim \lambda/2$, hence, in practice, $L \gtrsim 0.7\lambda$). ***Rule 2:*** the advantage requires an intermediate $|\varepsilon|$; as $|\varepsilon| \to \infty$ (seawater, PEC), $s_B \to 0$ and the TM route is no longer suppressed, so straight monopoles of 0.25–0.6λ then exceed a TE-dominated element. ***Rule 3:*** curvature of quarter-wave elements is ineffective (Proposition 3(iii)). ***Rule 4:*** in arrays, elements should share the mast and arc circle (Proposition 4), and the TE lobes ($\phi = \pm 90°$) should be directed into the coverage sector by the parasitic elements, since the TE radiation of a single element is bidirectional (Proposition 3(i)).

## III. METHODOLOGY

The closed-form analysis of Section II sets the design rules and the search ranges. For the single element, BO on an analytical current model locates the optimal geometry, which HFSS then refines for match and efficiency. For the arrays, classical Yagi–Uda rules and the common-arc rule initialize staged HFSS sweeps of the reflector, the driven element, and the director. The resulting sweep design is used to validate a fast thin-wire MoM surrogate, on which global, trust-region, and impedance-aware BO explore seven variables; the candidates on the resulting Pareto front are verified in HFSS before a final design is chosen. Finally, the currents of all final designs are extracted from the full-wave solutions, propagated to the far field with the Fresnel reflection of (1)–(6), and validated against independent references.

### A. FULL-WAVE SITE MODEL

The site was modeled in ANSYS HFSS® at 15 MHz ($\lambda = 19.99$ m), as shown in Fig. 2(c). Each element is a 5-cm-radius copper tube; the driven element is fed by a lumped port across a 10-cm gap, and the parasitic elements are shorted to the screen. The elements stand on a 16.3 m × 16.3 m (0.82λ) PEC screen lying on a soil block 1.5 skin depths deep, split into a finely meshed near region (λ × λ around the feed) and a far region. The block is terminated by an impedance boundary $\eta_0/\sqrt{\varepsilon}$, and radiation boundaries lie λ/2 beyond the screen edge and λ/4 above the antenna. The reference soil is moist ground ($\varepsilon_r = 10$, $\sigma = 10$ mS/m, $|\varepsilon| = 15.6$, loss tangent 1.2); the comparison uses the representative ground constants of Table 1, spanning seawater to very dry land [26]. Refining the near-soil mesh to λ/20 changed $Z_{\text{in}}$ by less than 0.5%. Over 400 full-wave variations were solved in total, including 182 single-element variations, about 120 array-sweep variations, 30 robustness variations, and dedicated validation runs.

### B. CURRENT EXTRACTION

To remove the sinusoidal-current assumption of the analytical model, the current on every element is extracted from the full-wave solution. Two non-model polylines are drawn parallel to each element at distances $\pm r_0$ in the $y$ direction, i.e., normal to the element plane, so that they remain exactly $r_0$ from the wire axis along both the tube and the arc. Sampling the complex magnetic field $\mathbf{H}_\pm(\ell)$ at 201 points on each line, Ampère's law gives

$$I(\ell) \approx \pi r_0 \left[\boldsymbol{H}_+(\ell) \cdot \hat{\boldsymbol{\phi}}_+ + \boldsymbol{H}_-(\ell) \cdot \hat{\boldsymbol{\phi}}_-\right], \\ \hat{\boldsymbol{\phi}}_\pm = \pm\hat{\boldsymbol{t}} \times \hat{\boldsymbol{y}} \tag{15}$$

Averaging the two sides cancels the field components that are common to both lines, such as the fields of neighboring elements and of the arc curvature. For the single curved element, a common $H_y$ component of about 0.23 A/m on the arc was cancelled in this way. The offset $r_0$ must be large enough for the mesh to resolve the field and small compared with the arc radius and the element spacing. Section VI-A shows that the extraction converges for $r_0 = 30$–60 cm, and $r_0 = 60$ cm was used for all designs.

### C. FRESNEL FAR-FIELD POST-PROCESSING AND GAIN NORMALIZATION

HFSS's far-field transformation assumes free space outside the solution box, whereas the low-angle reflection forms far beyond it. The far field over soil is therefore computed from (1) and (2), generalized to arrays by the phase factors of Section II-C, using the extracted currents on the exact geometry, including the feed-gap segment. This is the reflection-coefficient approach of [16], here driven by full-wave currents and a full-wave efficiency. The gain is normalized as

$$G(\psi, \phi) = \eta \, \frac{4\pi r^2 \left(|E_\theta|^2 + |E_\phi|^2\right)}{2\eta_0 P_r} \tag{16}$$

where $P_r = (r^2/2\eta_0)\int_{2\pi} |\mathbf{E}^{\text{PEC}}|^2 \, d\Omega$ is the power that the same currents radiate over a PEC ground, so that the absolute current level cancels. The near-field efficiency $\eta$ is taken from the power balance of Section III-D. The realized gain is

$$G_R(\psi,\phi) = (1-|S_{11}|^2)\frac{G(\psi,\phi)}{L_{net}} \tag{17}$$

where $L_{\text{net}}$ is the loss of the matching network, if any. Low-reactance designs are fed directly from 50 Ω; low-resistance designs use a lossy L-network (series inductor of quality factor 150, shunt capacitor) designed from the fine impedance sweep by $Q_L = \sqrt{50/R_{\text{in}} - 1}$, $X_s = Q_L R_{\text{in}} - X_{\text{in}}$, and $X_p = -50/Q_L$ . No sinusoidal-current assumption remains in the results of Sections V–VII.

### D. NEAR-FIELD EFFICIENCY BY POWER BALANCE

The near-field efficiency is obtained from a power balance of the full-wave solution,

$$P_{acc} = P_{near} + P_{far} + P_{air} + P_{out}, \quad \eta = \frac{P_{air}}{P_{acc}} \tag{18}$$

where $P_{\text{near}}$ and $P_{\text{far}}$ are the volume integrals of $1/2\,\sigma|\mathbf{E}|^2$ over the two soil regions, $P_{\text{air}}$ and $P_{\text{out}}$ are the Poynting fluxes through the air and soil boundaries, and $P_{\text{acc}}$ is the accepted power. Only $P_{\text{air}}$ reaches the sky; the power that leaves through the soil boundaries travels inside the lossy ground. HFSS's native radiation efficiency counts both fluxes as radiated and is therefore not used. All balances were within −0.7% to +3.6%.

### E. VALIDATION PROTOCOL

Three independent checks were applied to every design. First, the extracted feed current was compared with $\sqrt{2P_{\text{acc}}/R_{\text{in}}}$ from the port quantities. Second, for four arrays, a separate HFSS run over an infinite PEC ground, where HFSS's far field is exact, was compared with the image-theory field of the currents extracted from that run. Third, each PEC run was checked by integrating its gain over the upper hemisphere, which must give $\int G\, d\Omega/4\pi \approx 1$ for a lossless antenna. This check detected one run in which only 60% of the accepted power appeared in the far field because the ground was not continuous beyond the screen; its pattern shape and F/B still agreed with the extracted currents, so only the run, not the method, was at fault.

### F. OPTIMIZATION FRAMEWORK

#### 1) SINGLE ELEMENT: BAYESIAN OPTIMIZATION OF THE ANALYTICAL MODEL

For the single element, the design variables are $L$, the straight fraction $r = h_t/L$, and $R$, subject to $\theta_c = (1-r)L/R \le 90°$. The pattern-only metric $\bar{G}_s$ ($\eta = 1$, ideal match) of the analytical model, which evaluates (1)–(7) with a sinusoidal current in about 1 s, was maximized by BO over $L \in [0.2\lambda, 0.9\lambda]$, $r \in [0.3,1]$, and $R \in [0.02\lambda, 0.35\lambda]$. A Gaussian-process surrogate with a Matérn-5/2 kernel and maximum-likelihood hyperparameters [22] was initialized with 10 random samples and updated over 40 iterations by the expected-improvement (EI) acquisition [20],

$$EI(\boldsymbol{x}) = (\mu(\boldsymbol{x}) - J^+)\Phi(z) + \sigma(\boldsymbol{x})\varphi(z), \quad z = \frac{\mu(\boldsymbol{x}) - J^+}{\sigma(\boldsymbol{x})} \tag{19}$$

where $\mu$ and $\sigma$ are the surrogate mean and standard deviation, $J^+$ is the best observed objective, and $\Phi$ and $\varphi$ are the standard normal distribution and density. Geometries with $\theta_c > 90°$ were mapped to the nearest feasible arc. For a fair comparison, BO and random search started from the *same* 10 initial samples in each of three independent runs. HFSS then refined the region found by BO with a parametric grid and broader length sweeps and supplied $Z_{\text{in}}$ and $\eta$.

#### 2) ARRAYS: KNOWLEDGE-BASED INITIALIZATION AND CONSTRUCTION RULE

The array search was initialized with the classical design rules for Yagi–Uda arrays [19]: a reflector about 3%–4% longer and a director about 5% shorter than the driven element, with spacings of about 0.2λ. Because the present elements are thinner ($d/\lambda = 0.005$) than those of [19] and are ground-mounted, no boom correction is needed. Following Proposition 4, all elements share the driven element's tube height $h_t$ and arc circle of radius $R$, and a nominal factor $k_n$ sets the arc angle $\theta_n = (k_n L - h_t)/(k_n R)$. The physical conductor length is then

$$L_n = h_t + R\theta_n = L + h_t\left(1 - \frac{1}{k_n}\right), \quad p_n = \frac{L_n}{L} = 1 + r\left(1 - \frac{1}{k_n}\right) \tag{20}$$

so that a nominal $k_n$ changes the physical length much less than $k_n$ itself; for $r = 0.25$, $k_r = 1.18$ gives $p_r = 1.038$. This *common-arc* rule was compared with an *exact-copy* rule in which every dimension of the parasitic element is scaled by $k_n$ (Section V-A).

#### 3) ARRAYS: FULL-WAVE PARAMETRIC SWEEPS

Each array was optimized by staged HFSS sweeps with far-field quantities evaluated during the solution (forward and backward gain at $\psi = 3°$, 4°, 10°, 20°, and 30°, together with $S_{11}$ and $Z_{\text{in}}$ at 15 MHz), which avoids storing fields and keeps each variation fast. The reflector factor and spacing were swept first, then the driven length was retuned, and finally the director factor and spacing were swept. The same procedure, budget, and selection rule were applied to the straight baselines.

#### 4) ARRAYS: THIN-WIRE MOM SURROGATE AND BAYESIAN OPTIMIZATION

The 3-element array has seven continuous variables, $\mathbf{x} = (L, r, \theta_c, p_r, p_d, s_r, s_d)$, which is beyond a dense parametric grid. A fast and physically faithful objective was built from a thin-wire Galerkin MoM model [27] of the array and its image in the ground plane, with triangular basis functions $\Lambda_m$ on 0.25-m segments, eight-point Gauss quadrature, and the reduced kernel:

$$Z_{mn}=\frac{j\omega\mu_0}{4\pi}\iint \hat{\boldsymbol{t}}_m\cdot\hat{\boldsymbol{t}}_n\Lambda_m\Lambda_n g\,d\ell\,d\ell' -\frac{j}{4\pi\omega\varepsilon_0}\iint \Lambda_m'\Lambda_n' g\,d\ell\,d\ell' \quad (21)$$

A delta-gap source at the driven element's ground junction gives the currents of all elements; their far field over moist soil then follows from (1), (2), and (16), and the monopole input impedance from half the dipole impedance. Here $g=e^{-jkR_a}/R_a$ with $R_a=(|\mathbf{r}-\mathbf{r}'|^2+a^2)^{1/2}$ and wire radius $a$. One evaluation takes about 1.5 s. The model was validated against the extracted HFSS currents of the sweep-optimized 3-element array (Section V-D) before use. The objective penalizes a low F/B and a low input resistance:

$$J(\boldsymbol{x})=\bar{G}_f-max\left(0,\,12-\overline{F/B}\right)-\beta max(0,\,R_0-R_{in}) \quad (22)$$

where $\bar{G}_f$ is the pattern-only forward-sector gain over moist soil, $\overline{\mathrm{F/B}}$ is the mean F/B over $\psi=3°$–$30°$ in dB, and $(\beta,R_0)$ is $(0.5{,}12\ \Omega)$ for the gain-oriented objective or $(0.15{,}25\ \Omega)$ for the impedance-aware objective. Infeasible geometries (any $\theta_n>1.65$ rad) receive a fixed low score. Three strategies were compared with equal budgets: global BO over the full box (three seeds, 12 initial samples plus 48 EI iterations), random search, and a trust-region BO [23] confined to ±15% of the normalized box around the sweep-optimized design (80 evaluations). The three candidates that define the resulting Pareto front were then simulated in HFSS, including current extraction and power balance.

#### 5) FAIRNESS PROTOCOL

To ensure a fair comparison with straight antennas, all straight baselines were optimized with the same staged sweeps, the same calculations, and the same selection rule (maximum forward-sector gain subject to F/B ≥ 10 dB and $|S_{11}|<-10$ dB, or a lossless match with an L-network). Physical, not nominal, length ratios were used for straight parasitic elements, because applying the curved array's nominal factors to straight elements would make the reflector 18% too long. In addition, straight arrays with element lengths and spacings *identical* to the curved arrays isolate the effect of curvature alone.

### G. DESIGN PROCEDURE

The complete procedure, which can be reused for other frequencies, grounds, and element counts, is summarized as Algorithm 1.

## IV. SINGLE CURVED ELEMENT

### A. OPTIMIZED DESIGN

Fig. 4 compares the best curved variant at each length with the straight element over moist soil. Up to 10 m, the two agree within 0.1 dB, as Propositions 1 and 3(iii) predict. Above 11 m they separate: the curved metric rises to +1.15 dBi at 15.5 m and then declines, whereas the straight element improves only slowly, to −0.77 dBi at 17 m. Three independent BO runs converged to $L=15.0$–$15.4$ m, $r=0.32$–$0.36$, and $R=6.3$–$6.5$ m with $\bar{G}_s=1.13$–$1.20$ dBi, whereas random search, started from the same initial samples and given the same budget of 50 evaluations, reached only 0.36–1.05 dBi (Fig. 5). Sampled HFSS variants with $L=14.75$–$15.5$ m, $r=0.20$–$0.35$, and $R=6.5$–$8.5$ m lie within 0.15 dB of the optimum. Within this plateau, the 15.5-m peak is mismatched ($|S_{11}|=-8.6$ dB), so the 15-m design ($h_t=3.75$ m, $R=7.5$ m, $\theta_c=86°$, total height 11.2 m) was selected: it gives up only 0.05 dB of pattern gain and is nearly resonant.

**Algorithm 1** Curved Yagi–Uda design over lossy ground

1: From the target sector and ground constants, compute $s_B$ and $\delta$ with (8)–(12), and choose $L\gtrsim 0.7\lambda$ so that $\psi^*$ of (14) falls near the sector.
2: Optimize the single element $(L,r,R)$ by BO on the analytical model, then refine it with full-wave sweeps for match and efficiency.
3: Initialize the parasitic elements by the classical rules [19] with the common-arc rule (20); retune the driven length for resonance in the coupled array.
4: Sweep the reflector and then the director with far-field calculations evaluated during the solution, and select by maximum $\bar{G}_f$ subject to F/B and matching constraints.
5: Validate a MoM surrogate against the extracted full-wave currents and run global, trust-region, and impedance-aware BO; confirm the best candidates in the full-wave model.
6: Extract the currents (15), close the power balance (18), and compute the realized gain on all grounds with (16) and (17).
7: Check the post-processing over PEC and quantify robustness to construction errors.

**TABLE 1.** PATTERN-ONLY SECTOR GAIN $\bar{G}_s$ (DBI) OF SINGLE ELEMENTS VERSUS GROUND ($\eta=1$)

| Ground | $\varepsilon_r$, $\sigma$ (S/m) | $\lvert\varepsilon\rvert$ | $\psi_B$ (°) | λ/4 ref. | Straight 0.75λ | Best straight ($L$ [m]) | Curved 0.75λ |
|---|---|---|---|---|---|---|---|
| PEC | – | ∞ | – | 4.54 | −0.79 | 5.55 (11.5) | 1.58 |
| Seawater | 80, 5 | 5992 | 0.7 | 4.08 | −1.66 | 4.87 (11.5) | 1.43 |
| Wet | 20, 0.03 | 41.1 | 8.9 | 0.07 | −2.35 | 0.07 (4.67) | 1.06 |
| Average | 13, 0.005 | 14.3 | 14.9 | −1.74 | −0.85 | −0.15 (13.5) | 0.96 |
| Moist | 10, 0.01 | 15.6 | 14.4 | −1.73 | −0.92 | −0.77 (17) | 1.10 |
| Poor | 5, 0.001 | 5.14 | 23.9 | −3.57 | 0.53 | 1.22 (14) | 0.88 |
| Very dry | 3, 0.0001 | 3.00 | 30.0 | −4.27 | 1.06 | 1.87 (14) | 0.66 |

Best straight: highest value over $L=4.67$–$17$ m on each ground, with its length in parentheses. λ/4 ref.: 0.234λ resonant monopole.

**TABLE 2.** FINAL SINGLE-ELEMENT DESIGNS ON MOIST SOIL (15 MHZ, 50-Ω FEED)

| Quantity | λ/4 ref. | Straight 0.75λ | Curved 0.75λ |
|---|---|---|---|
| $L$ / height (m) | 4.67 / 4.67 | 15.0 / 15.0 | 15.0 / 11.2 |
| $Z_{\mathrm{in}}$ (Ω) | 34.3 + j3.8 | 61.8 + j27.1 | 53.7 + j4.3 |
| $\lvert S_{11}\rvert$ (dB) | −14.3 | −11.8 | −25.3 |
| −10-dB bandwidth (%) | 11.97 | 5.96 | 6.0 |
| $\eta$ (power balance) | 0.754 | 0.896 | 0.930 |
| $\bar{G}_s$, 3°–30° (dBi) | −3.12 | −1.69 | 0.77 |
| $\bar{G}$, 3°–10° (dBi) | −6.69 | −4.90 | −4.55 |
| $G_s(\phi=90°)$ (dBi) | −3.12 | −1.69 | 3.57 |

Realized, solid-angle-averaged gains (17).

Table 2 summarizes the final full-wave runs. With equal conductor length, curvature lowers the mast by 25%, improves the match from −11.8 to −25.3 dB, raises efficiency $\eta$ from 90% to 93%, and increases the realized azimuth-averaged sector gain by 2.5 dB, with no penalty at 3°–10°. Its −10-dB bandwidth (6.0%, 14.50–15.40 MHz from a 25-kHz sweep) is comparable to that of the straight element (5.96%). The quarter-wave reference loses 25% of its power in the soil, so 0.9 dB of its 3.9-dB deficit is due to efficiency rather than pattern.

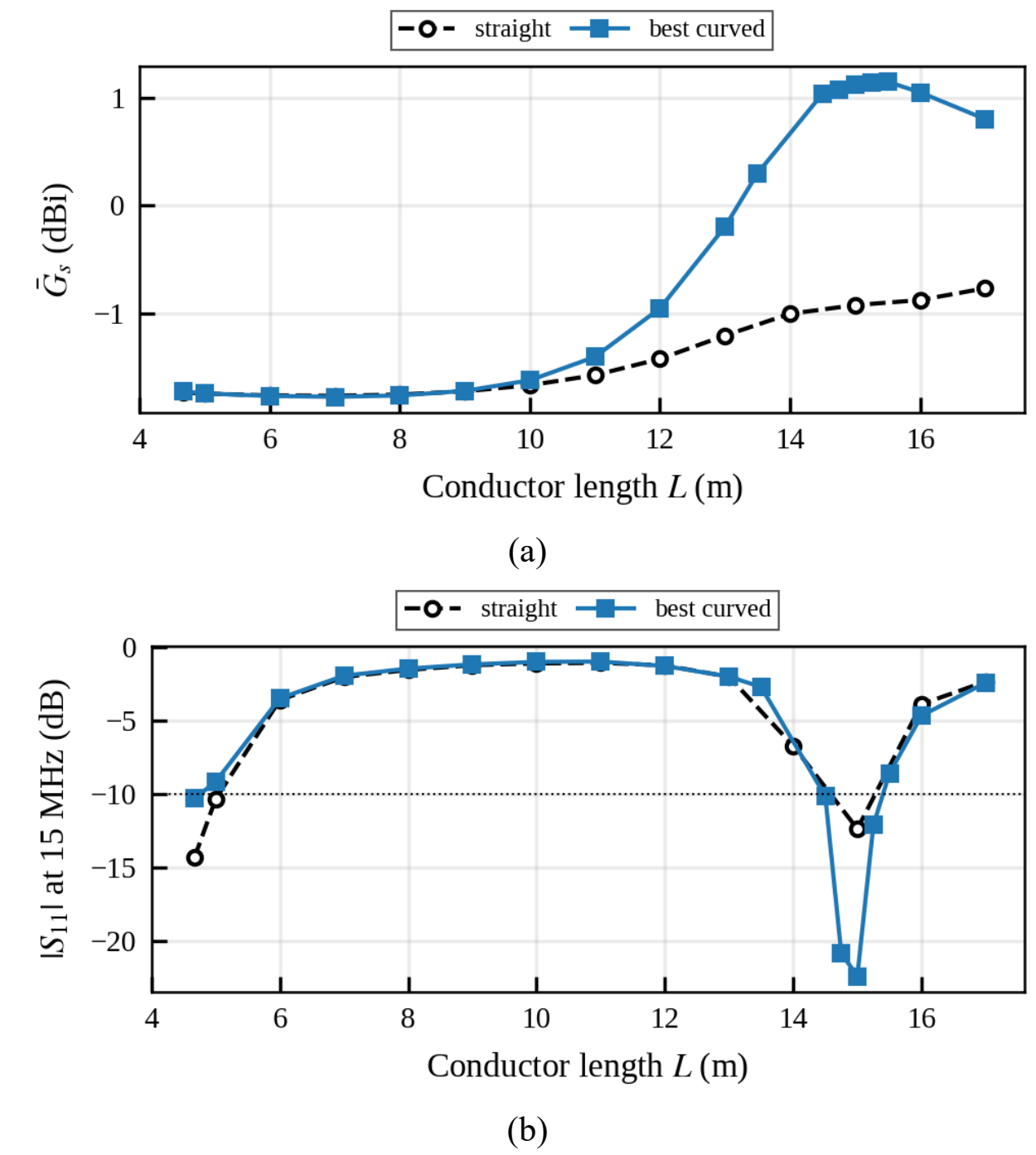

**FIGURE 4.** Single element versus conductor length $L$ on moist soil (HFSS): (a) pattern-only sector gain $\bar{G}_s$ of the straight element and of the best curved variant at each length; (b) $|S_{11}|$ at 15 MHz (50-Ω feed).

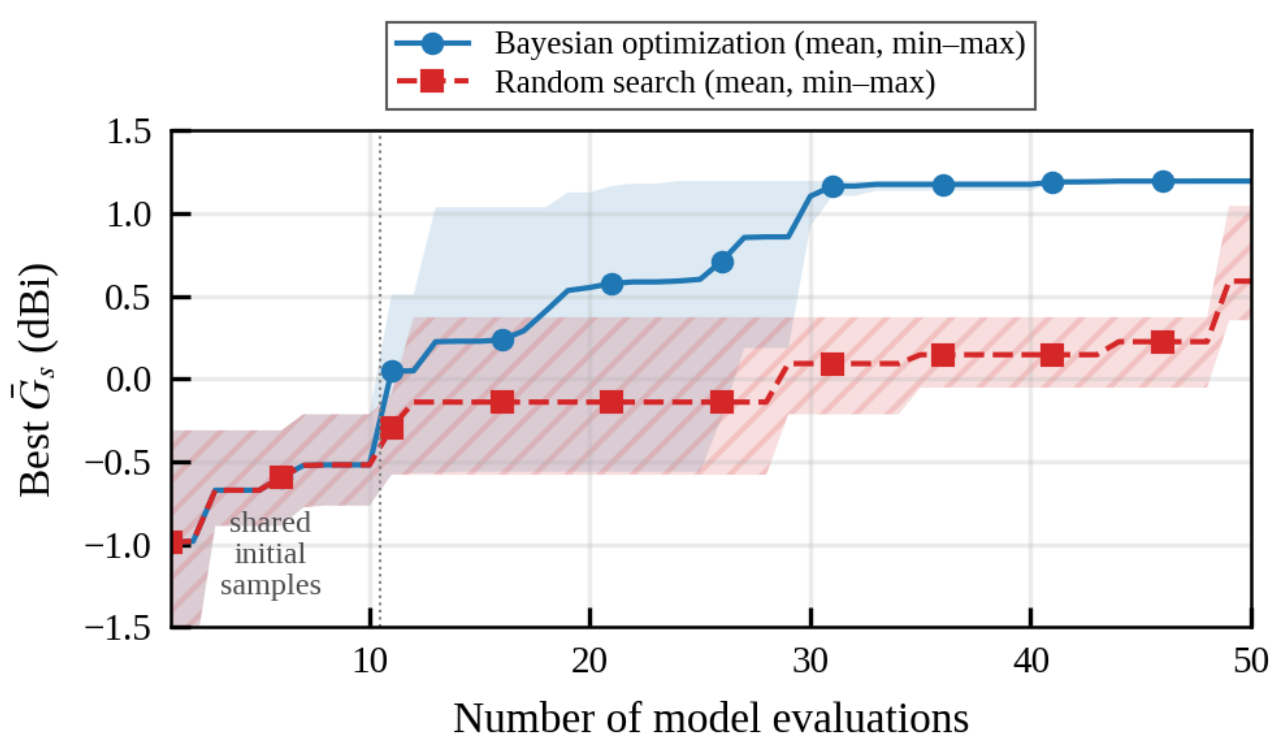

**FIGURE 5.** Convergence of BO and random search on the analytical single-element model over moist soil: mean (lines) and min–max range (shading) of three runs; both methods start from the same 10 initial samples (dotted line).

## B. MECHANISM AND SOIL DEPENDENCE

Two effects are superimposed. First, a pattern effect: the 15-m straight element is a poor low-angle radiator even over PEC (−0.79 dBi), because its main lobe lies near 45°, and curvature redirects its radiation into a broadside TE lobe. Second, a soil effect: from PEC to moist soil, the TM sector power of the quarter-wave reference falls from 2.85 to 0.67 (−6.3 dB), whereas the TE sector power of the curved element falls only from 1.15 to 1.03 (−0.5 dB), in line with Propositions 1 and 2. TE radiation thus carries 80% of the curved element's sector power and peaks broadside (3.6 dBi at $\phi = 90°$), as Proposition 3(i) predicts; its weighted arc height $\bar{z} = 9.45$ m places $\psi^*$ at 32° by (14) (Fig. 6). For the 4.67-m element bent by up to 67°, including the geometry of [15], the TE share stays below 0.5% and $\bar{G}_s$ changes by less than 0.04 dB, consistent with Proposition 3(iii).

Table 1 separates the two effects and defines the advantage window. The 0.575λ straight monopole, optimal over seawater and PEC, loses 5.9–7.1 dB on land, whereas the curved element loses at most 0.9 dB; on every land ground, the curved element therefore exceeds it by 1.0–2.6 dB. Even with the straight length re-optimized for each soil, the curved element remains superior on wet, average, and moist soils (+1.0 to +1.9 dB). Over seawater and PEC, the 0.575λ straight monopole remains better by 3.4–4.0 dB, consistent with Rule 2. The single element is, however, bidirectional broadside (−7.5 to −5.9 dBi in the arc plane), which motivates the arrays of Section V.

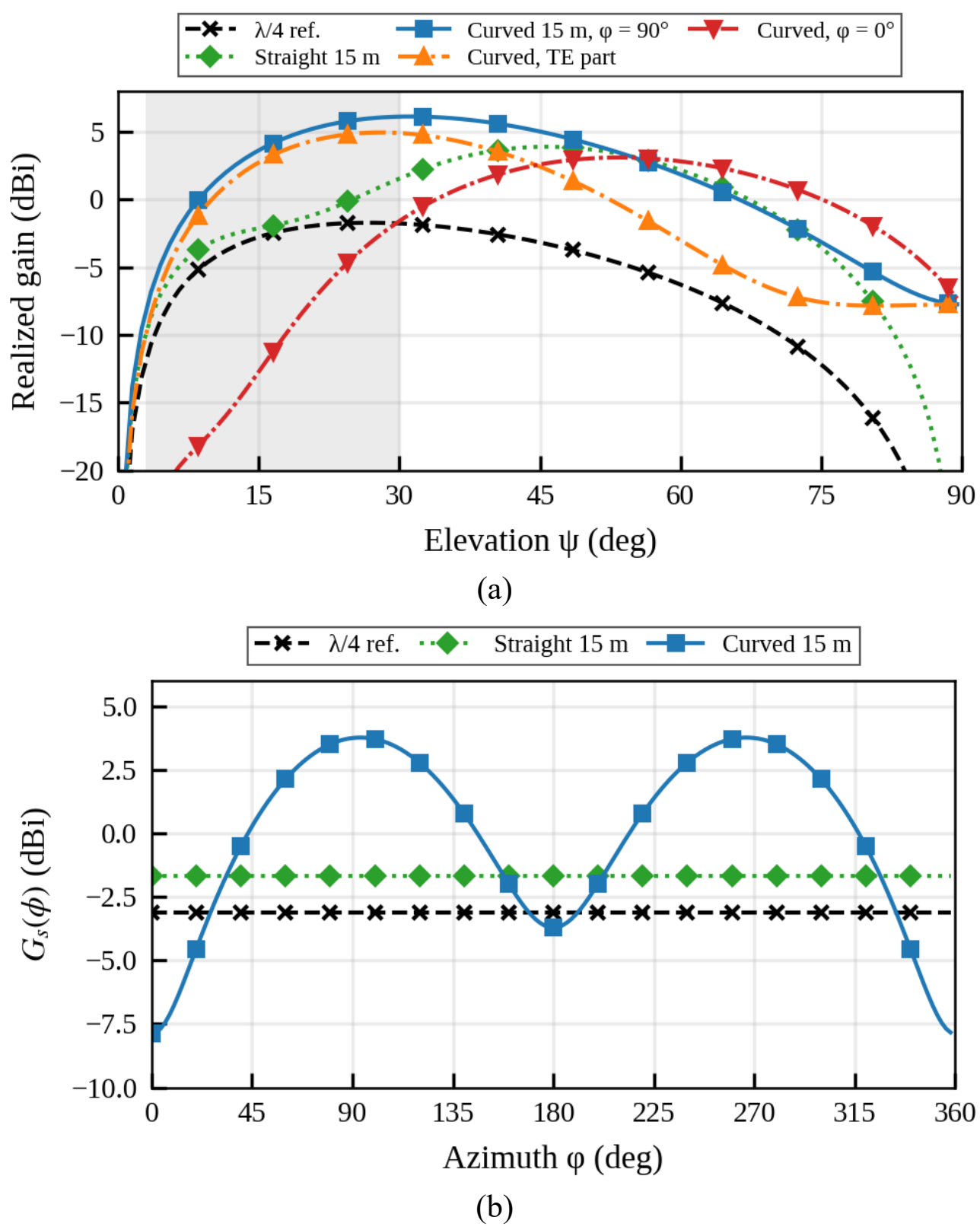

**FIGURE 6.** Single element over moist soil (realized): (a) elevation cuts, including the curved element's broadside TE part and its arc-plane cut; (b) 3°–30° sector gain $G_s(\phi)$ versus azimuth, showing the two broadside TE lobes of the curved element.

## V. CURVED YAGI–UDA ARRAYS: DESIGN AND OPTIMIZATION

### A. CONSTRUCTION RULE

Two construction rules were compared for a 3-element array. With the common-arc rule of (20), a reflector with $k_r = 1.10$ at $s_r = 5$ m and a director with $k_d = 0.93$ at $s_d = 4$ m gave a native HFSS forward-sector gain of 7.82 dBi and an F/B of 12.4–17.0 dB. With the exact-copy rule and comparable nominal factors ($k_r = 1.22$, $s_r = 4.5$ m, $k_d = 0.90$, $s_d = 4$ m), the forward gain fell to 5.35 dBi and the F/B to 5.7–5.9 dB, and the driven resonance moved to 15.31 MHz. The reason is the meaning of $k$: under the exact-copy rule, $k_r = 1.22$ makes the reflector 22% longer, far from the 3%–6% that the Yagi–Uda action requires, whereas under the common-arc rule the same factor lengthens it by only 4.1% by (20). The common-arc rule is also mechanically simple, because all elements use identical masts and the same arc former, and it satisfies the assumption of Proposition 4. It was therefore adopted for all curved arrays. Table 3 lists the final geometries.

**TABLE 3.** FINAL ARRAY GEOMETRIES (COMMON-ARC RULE FOR CURVED ARRAYS)

| Design | Driven $L$ (m), $r$, $\theta_c$ | Reflector $L_r$ (m) $[p_r]$, $s_r$ (m) | Director $L_d$ (m) $[p_d]$, $s_d$ (m) | Height (m) |
|---|---|---|---|---|
| Curved 2-el | 14.50, 0.25, 85.9° | 15.05 [1.038], 4.50 | – | 11.0 |
| Curved 3-el (sweep) | 14.50, 0.25, 85.9° | 15.05 [1.038], 4.50 | 14.35 [0.990], 4.00 | 11.0 |
| Curved 3-el (BO) | 14.57, 0.272, 88.2° | 15.11 [1.037], 4.94 | 14.25 [0.978], 4.40 | 11.0 |
| Straight λ/4 2-el | 4.54, 1, – | 4.90 [1.08], 4.00 | – | 4.9 |
| Straight λ/4 3-el | 4.54, 1, – | 4.90 [1.08], 4.00 | 4.36 [0.96], 4.00 | 4.9 |
| Straight, same length (3-el) | 14.50, 1, – | 15.05 [1.038], 4.50 | 14.35 [0.990], 4.00 | 15.1 |

$p$: physical length ratio to the driven element. All curved elements share $h_t$ and $R$ of the driven element.

### B. TWO-ELEMENT ARRAY

Starting from the rules of [19] with $L = 15$ m, the reflector first detuned the driven element: the coupled array resonated below 15 MHz. Shortening the driven element to 14.5 m restored the resonance; alone, the 14.5-m element resonates at 15.5 MHz ($|S_{11}| = -10.3$ dB at 15 MHz), so the mutual coupling pulls the array resonance down by about 0.5 MHz. The reflector was then swept over $k_r = 1.03$–$1.30$ ($p_r = 1.007$–$1.058$) and $s_r = 3$–$6$ m in 48 S-parameter and 37 far-field variations (Fig. 7 shows the three spacings sampled on the complete grid). The spacing mainly controls the match: $s_r \leq 3$ m never reaches −10 dB, whereas $s_r \geq 4.5$ m is self-matched with about 5%–7% bandwidth for every reflector length. The reflector length mainly controls the backward lobe: the mean F/B rises from about 6 dB at $p_r = 1.012$ to 12–18 dB at $p_r = 1.038$–$1.058$, while the forward gain declines only slowly beyond $p_r \approx 1.04$. Table 4 lists four candidates; the balanced candidate 2B ($k_r = 1.18$, $p_r = 1.038$, $s_r = 4.5\text{ m} = 0.225\lambda$) was selected.

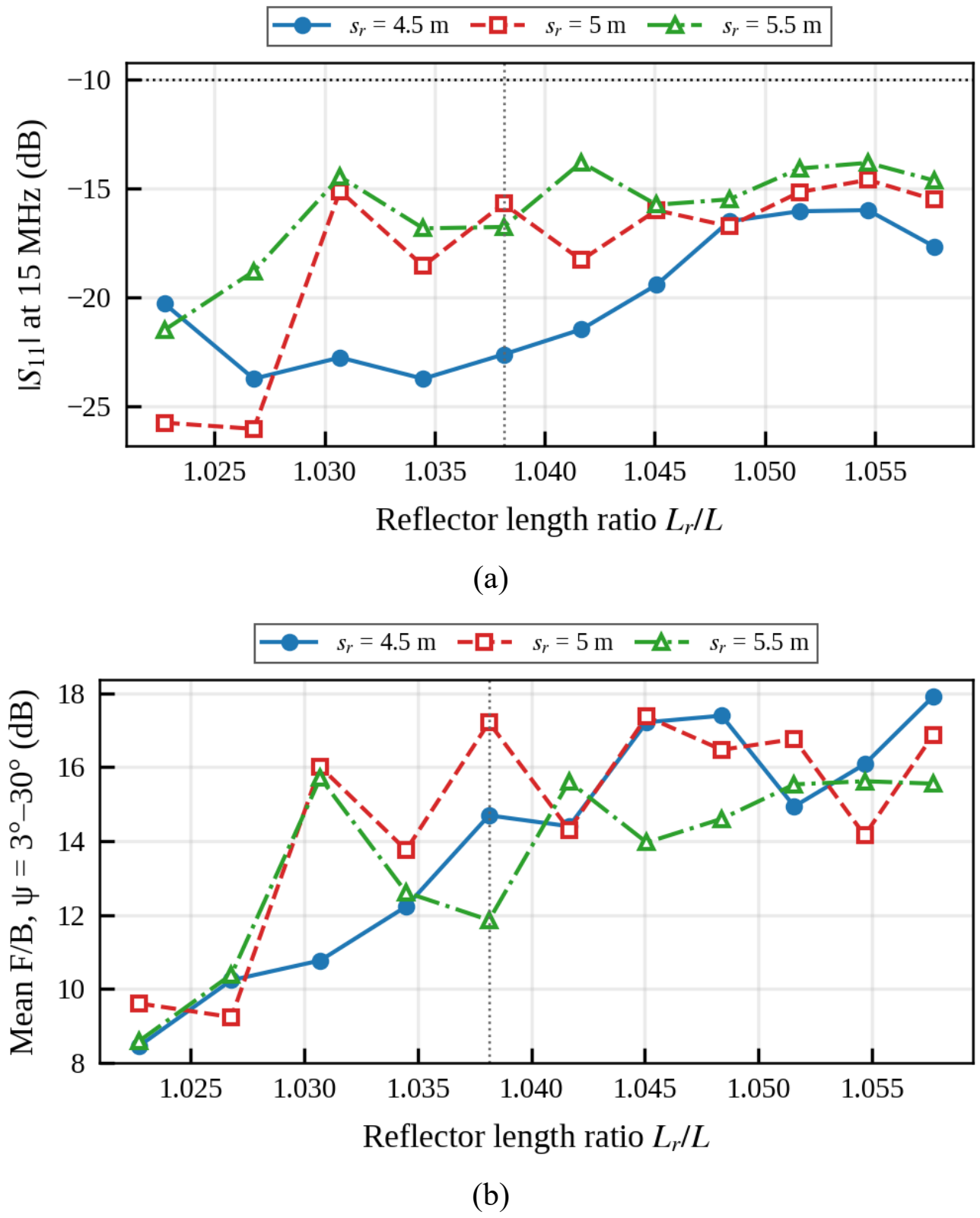

**FIGURE 7.** Two-element curved array sweep (HFSS, moist soil, driven $L = 14.5$ m): (a) $|S_{11}|$ at 15 MHz and (b) mean F/B over the 3°–30° sector versus the physical reflector length ratio for $s_r = 4.5$, 5.0, and 5.5 m on the complete grid $k_r = 1.10$–$1.30$ (step 0.02). The dotted line marks the selected design 2B.

**TABLE 4.** TWO-ELEMENT CANDIDATES (HFSS NATIVE, $\phi = 90°$)

| Option | $k_r$ $[p_r]$ | $s_r$ (m) | $G$ at 30°/20°/10°/4° (dBi) | F/B (dB) | $\lvert S_{11}\rvert$ (dB) |
|---|---|---|---|---|---|
| 2A, max. F/B | 1.22 [1.045] | 4.5 | 9.8 / 7.8 / 4.8 / 2.7 | 15.6–18.6 | −19.4 |
| 2B, balanced | 1.18 [1.038] | 4.5 | 10.0 / 8.0 / 5.0 / 2.9 | 12.7–16.8 | −22.6 |
| 2C, max. gain | 1.16 [1.034] | 4.5 | 10.1 / 8.1 / 5.2 / 3.1 | 10.5–14.1 | −23.8 |
| 2D, best match | 1.10 [1.023] | 5.0 | 10.0 / 8.0 / 5.0 / 3.0 | 8.2–11.2 | −25.8 |

Relative to the 14.5-m element alone (native forward-sector gain 2.76 dBi), the reflector adds 3.6 dB. The final 2B array has $Z_{\text{in}} = 57.4 + j3.4$ Ω, $|S_{11}| = -22.4$ dB, a −10-dB bandwidth of 6.8% (14.64–15.66 MHz), and $\eta = 0.947$.

### C. THREE-ELEMENT ARRAY: DIRECTOR SWEEP

With the 2B reflector fixed, the director was swept over $k_d = 0.718$–$0.98$, i.e., physical ratios $p_d = 0.90$–$0.995$ by (20), and $s_d = 3$–$5$ m in 18 variations (Fig. 8, solid lines). The forward gain at high elevations is almost independent of the director (11.0–11.4 dBi at 30°); the director acts mainly on the low angles, where longer directors add up to 0.35 dB at 3°–10°. As in classical Yagi–Uda arrays, the director trades gain against F/B and input resistance: shorter directors give an F/B of 17–19 dB and $R_{\text{in}}$ up to 25 Ω, longer directors more low-angle gain but lower F/B and $R_{\text{in}}$, and $s_d = 3$ m is uniformly worse. Short directors ($p_d = 0.90$–$0.94$) form a distinct regime: the array becomes self-matched ($|S_{11}| = -13.5$ to $-21.7$ dB, $R_{in} = 33$–$45\ \Omega$) with an F/B of 17–22 dB, but its gain at 10° is 0.9–1.8 dB lower than with near-resonant directors. By the selection rule, $k_d = 0.96$ and $s_d = 4$ m ($p_d = 0.990$) were chosen. A slightly longer director ($k_d = 0.98$) raises the low-angle gain by a further 0.2 dB but lowers the F/B to 11.6–12.8 dB and $R_{in}$ to 11.4 Ω, which narrows the matched bandwidth below 1%; $k_d = 0.96$ was therefore retained. This array needs an L-network: its fine-sweep impedance $13.8 - j0.2\ \Omega$ is matched by 0.24 µH and 343 pF, which gives a matched −10-dB bandwidth of only 1.06% because its quality factor, estimated from the reactance slope as $Q \approx (f/2R_{in})\, dX_{in}/df$ [28], is about 62.

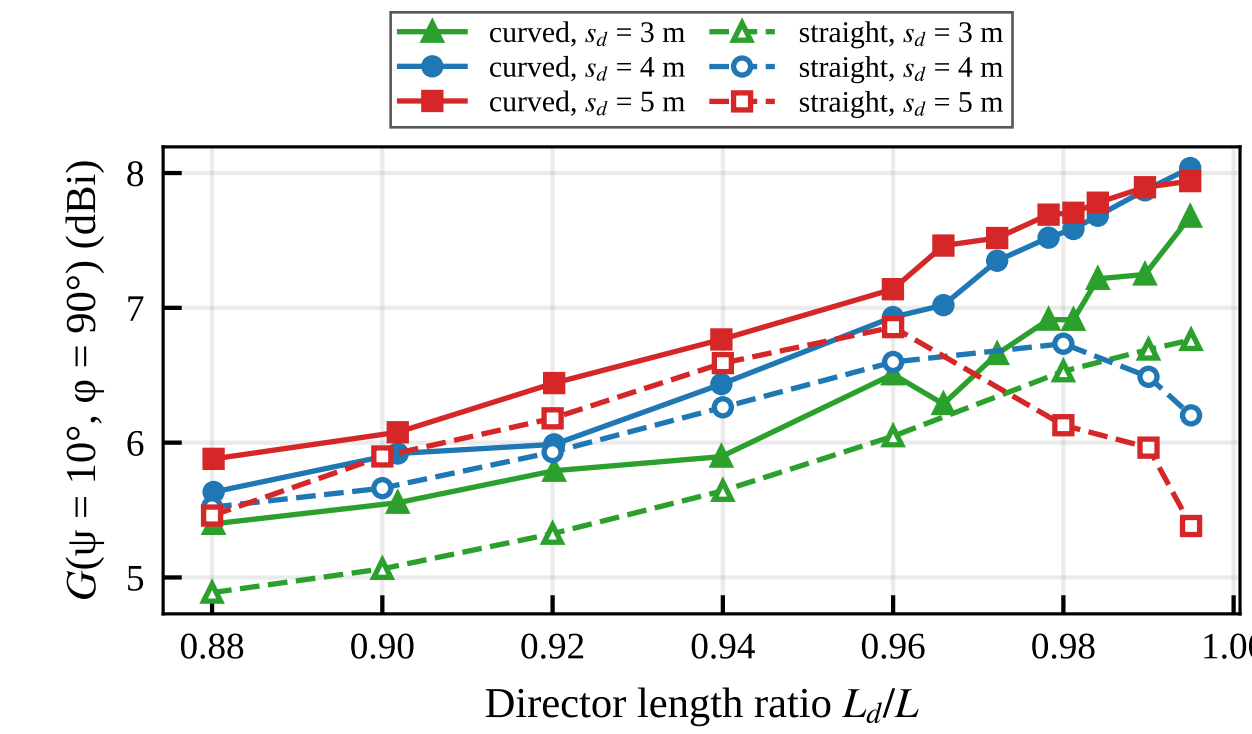

(a)

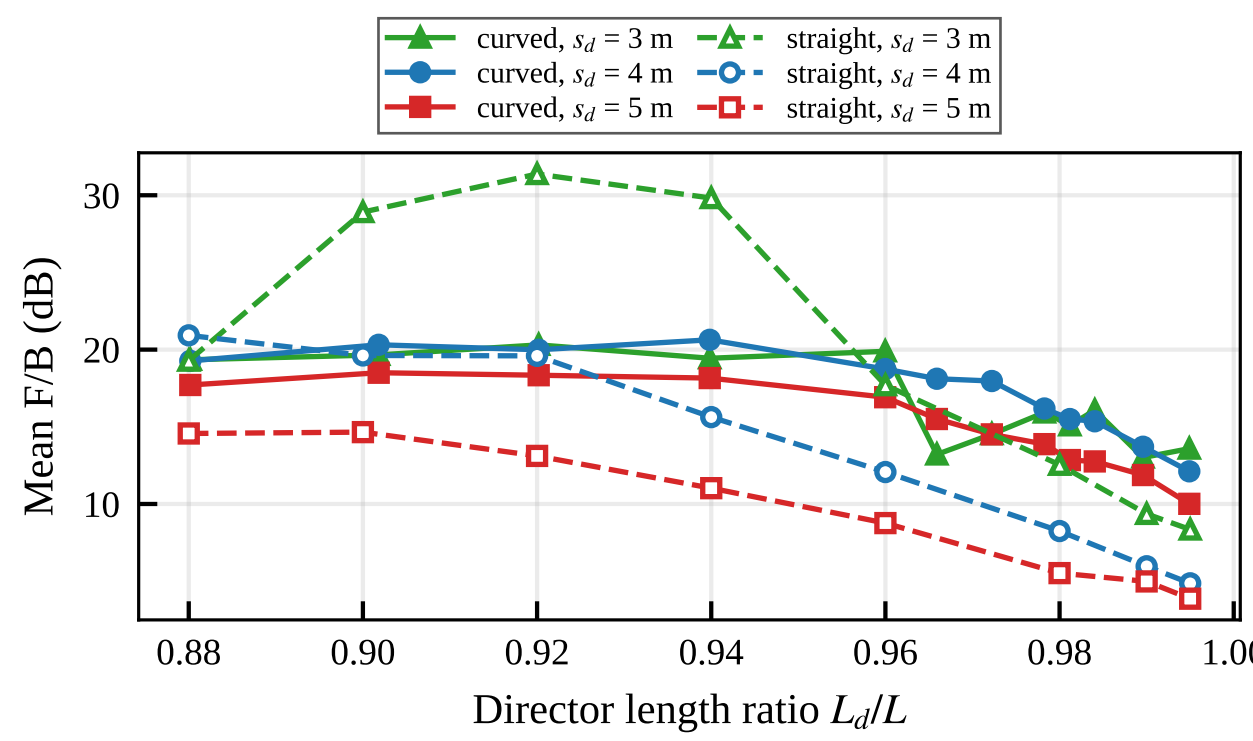

(b)

**FIGURE 8.** Director sweeps (HFSS, moist soil): (a) gain at $\psi = 10°$, $\phi = 90°$ and (b) mean F/B versus the physical director length ratio for the curved array (solid lines, filled markers) and the straight λ/4 array (dashed lines, open markers) at three spacings $s_d$. For the curved array, the nominal factors $k_d = 0.718$–$0.98$ correspond by (20) to $p_d = 0.90$–$0.995$.

### *D. THREE-ELEMENT ARRAY: BAYESIAN OPTIMIZATION*

Before the MoM surrogate was used, it was checked against the extracted HFSS currents of the sweep-optimized array over PEC. It gives $R_{in} = 13.8\ \Omega$ (HFSS: 13.6 Ω), currents on all three elements within a few percent in magnitude and 5° in phase, and a pattern-only forward-sector gain over moist soil of 8.87 dBi, against 8.85 dBi from the HFSS currents. Its reactance is offset by about 7 Ω because of the different feed-gap models, which is why the final candidates were confirmed in HFSS. Table 5 summarizes the optimization stages, and Table 6 and Fig. 9 the BO results. It is worth noting that over the full budget, BO and random search reach similar best values because the objective has a broad near-optimal plateau; BO, however, reaches 8.3 dBi in all runs (within 15–36 evaluations), whereas random search does so in one of three runs (after 47).

**TABLE 5.** OPTIMIZATION STAGES AND COMPUTATIONAL COST

| Stage | Model | Variables | Evaluations | Cost per evaluation | Outcome |
|---|---|---|---|---|---|
| Single-element BO | Analytical current | $L, r, R$ | 600 | ≈1 s | $L \approx 15$ m, $\bar{G}_s$ = 1.13–1.20 dBi |
| Single-element sweeps | HFSS site | $L, r, R$ | 182 | ≈1.5 min (no fields) | 15-m design, 0.15-dB plateau |
| 2-el sweeps | HFSS site | $k_r, s_r, L$ | ≈85 | ≈2 min (no fields) | 2B, self-matched |
| 3-el director sweep | HFSS site | $k_d, s_d$ | 33 | ≈4 min (no fields) | $k_d = 0.96$, $s_d$ = 4 m |
| 3-el global BO | MoM + Fresnel | $L, r, R, \theta_c, p_r, p_d, s_r, s_d$ | 600 | ≈1.5 s | 8.46–8.56 dBi (MoM) |
| 3-el trust-region BO | MoM + Fresnel | $L, r, R, \theta_c, p_r, p_d, s_r, s_d$ | 160 | ≈1.5 s | Pareto front, candidates A–C |
| Confirmation | HFSS site | – | 3 | ≈8 mins | A, B, C confirmed |

**TABLE 6.** BAYESIAN OPTIMIZATION OF THE 3-ELEMENT ARRAY: MOM PREDICTION AND HFSS CONFIRMATION (MOIST SOIL)

| Design | $\bar{G}_f$ MoM (dBi) | $\bar{G}_f$ HFSS (dBi) | $\bar{G}_f$ realized (dBi) | $R_{in}$ MoM / HFSS (Ω) | $Q$ | Matched BW (%) |
|---|---|---|---|---|---|---|
| C, gain end | 8.99 | 9.06 | 8.71 | 13.8 / 14.2 | 61 | 1.08 |
| Sweep-optimized | 8.87 | 8.85 | 8.51 | 13.8 / 13.8 | 62 | 1.06 |
| A, balanced (final) | 8.75 | 8.66 | 8.28 | 21.1 / 21.6 | 40 | 1.67 |
| B, impedance end | 8.44 | 8.43 | 8.11 | 25.0 / 25.0 | 31 | 2.14 |
| Global BO, best of 10 × 60 | 8.01–8.77 | – | – | 13.5–27.0 | – | – |
| Random search, best of 10 × 60 | 7.71–8.68 | – | – | 14.8–26.1 | – | – |

$\bar{G}_f$ MoM and HFSS: pattern only ($\eta = 1$); HFSS values from the Fresnel post-processing of extracted currents. Realized: with $\eta$, mismatch, and L-network loss (coil quality factor 150). F/B over 3°–30°: C 10.5–14.4, sweep 11.5–14.1, A 11.3–13.7, B 11.5–13.3 dB.

Three findings emerge. First, over 10 independent runs, global BO reaches a given gain with fewer evaluations than random search: after 30 and 40 evaluations its median gap to the best design is 0.76 and 0.63 dB, against 0.98 and 0.87 dB (p = 0.01 and 0.04), and 8 of 10 BO runs, against 5 of 10 random-search runs, come within 0.7 dB of the best design. With 60 evaluations, the difference narrows (p = 0.19) because the objective has a broad near-optimal plateau. Second, the trust-region BO, confined to ±15% around that design, finds a design that surpasses it. Candidate C ($L = 14.46$ m, $r = 0.276$, $\theta_c = 86.3°$, $k_r = 1.131$, $k_d = 0.972$, $s_r = 4.93$ m, $s_d = 4.10$ m) was predicted at 8.99 dBi and confirmed by HFSS at 9.06 dBi; its realized forward-sector gain over moist soil (8.71 dBi) exceeds that of the sweep-optimized array by 0.20 dB at the same input resistance (14.2 Ω), quality factor (61), and matched bandwidth (1.08%), at the cost of a lower F/B at the lowest angles (10.5 dB at 3°). Candidate C therefore dominates the sweep-optimized design, which does not lie on the Pareto front.

Third, and most useful in practice, the impedance-aware objective reveals a Pareto front between forward gain and input resistance (Fig. 9(b)) that the sweeps had not exposed: along the front, each additional ohm of input resistance costs only about 0.03 dB of gain up to $R_{\mathrm{in}} \approx 21\ \Omega$, and about 0.1 dB per ohm beyond it. The candidates at the balanced point (A: $L = 14.57$ m, $r = 0.272$, $\theta_c = 88.2^\circ$, $k_r = 1.157$, $k_d = 0.925$, $s_r = 4.94$ m, $s_d = 4.40$ m) and at the impedance end (B: $L = 14.48$ m, $r = 0.245$, $\theta_c = 84.4^\circ$, $k_r = 1.276$, $k_d = 0.918$, $s_r = 4.95$ m, $s_d = 4.45$ m) were also confirmed in HFSS (Table 6). Across all three candidates, the MoM surrogate predicted the pattern-only gain within 0.01–0.09 dB and the input resistance within 0%–3%. From C through A to B, the quality factor halves (61, 40, 31) and the matched bandwidth with an L-network doubles (1.08%, 1.67%, 2.14%) for 0.6 dB of realized gain. Candidate A, which widens the bandwidth of the sweep-optimized design by 58% at a cost of 0.23 dB, is adopted as the final 3-element design, because frequency agility and tolerance to construction errors (Section VIII-A) matter more for a radar than a few tenths of a decibel; candidates C and B are retained as the HFSS-confirmed ends of the front for applications that favor gain or bandwidth, respectively.

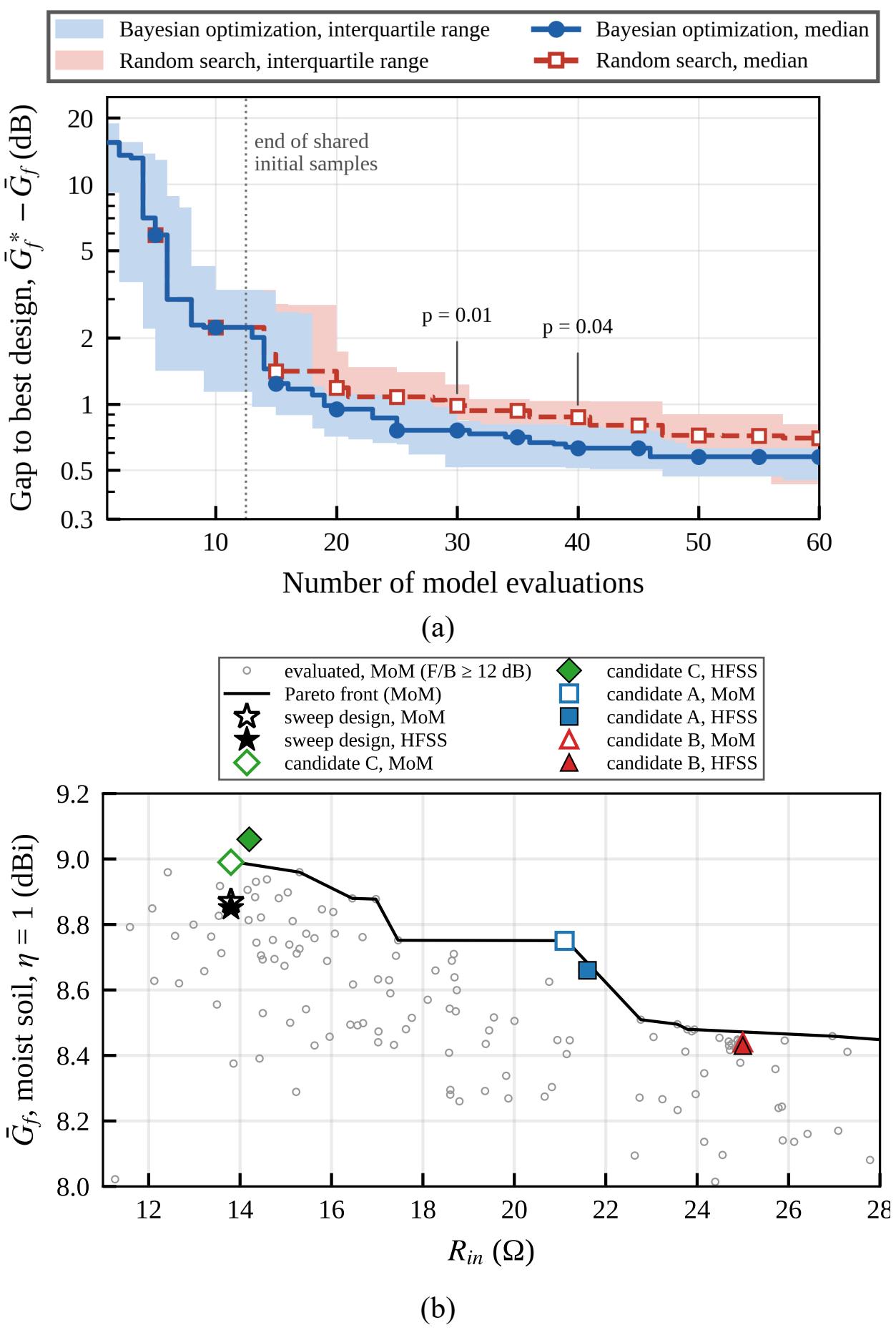

**FIGURE 9.** Bayesian optimization of the 3-element curved array (MoM surrogate): (a) Gap between the best forward-sector gain found so far and the best design found by any optimizer ($\bar{G}_f^* = 8.99$ dBi), for global BO and random search on the MoM surrogate: median (lines) and interquartile range (shading) of 10 runs with identical budgets and shared initial samples; p-values from a one-sided Mann–Whitney test; (b) pattern-only forward-sector gain versus input resistance of all evaluated designs with F/B ≥ 12 dB, the Pareto front of nondominated designs, the sweep-optimized design, and BO candidates A, B, and C; open markers: MoM prediction; filled markers: HFSS confirmation.

### E. CURRENTS AND PHASING

Fig. 10 shows the extracted currents of candidate A. The driven element exhibits the standing-wave form of the single element, with a current minimum near $\ell \approx 5$ m and a 180° phase reversal before the arc. The reflector carries about 0.5 times the driven current and leads it by about 137°; the director carries about 0.72 times the driven current and lags it by about 162°. These relative phases are nearly constant along both the tube and the arc, which confirms the assumption of Proposition 4 and shows that the parasitic elements steer the TE radiation of the arcs, not only the TM radiation of the tubes. The 2-element array behaves similarly (reflector 0.82–0.89 times the driven current, leading by 125°). In all curved arrays, the arc currents have uniform phase and one sign, which satisfies the conditions of Proposition 2.

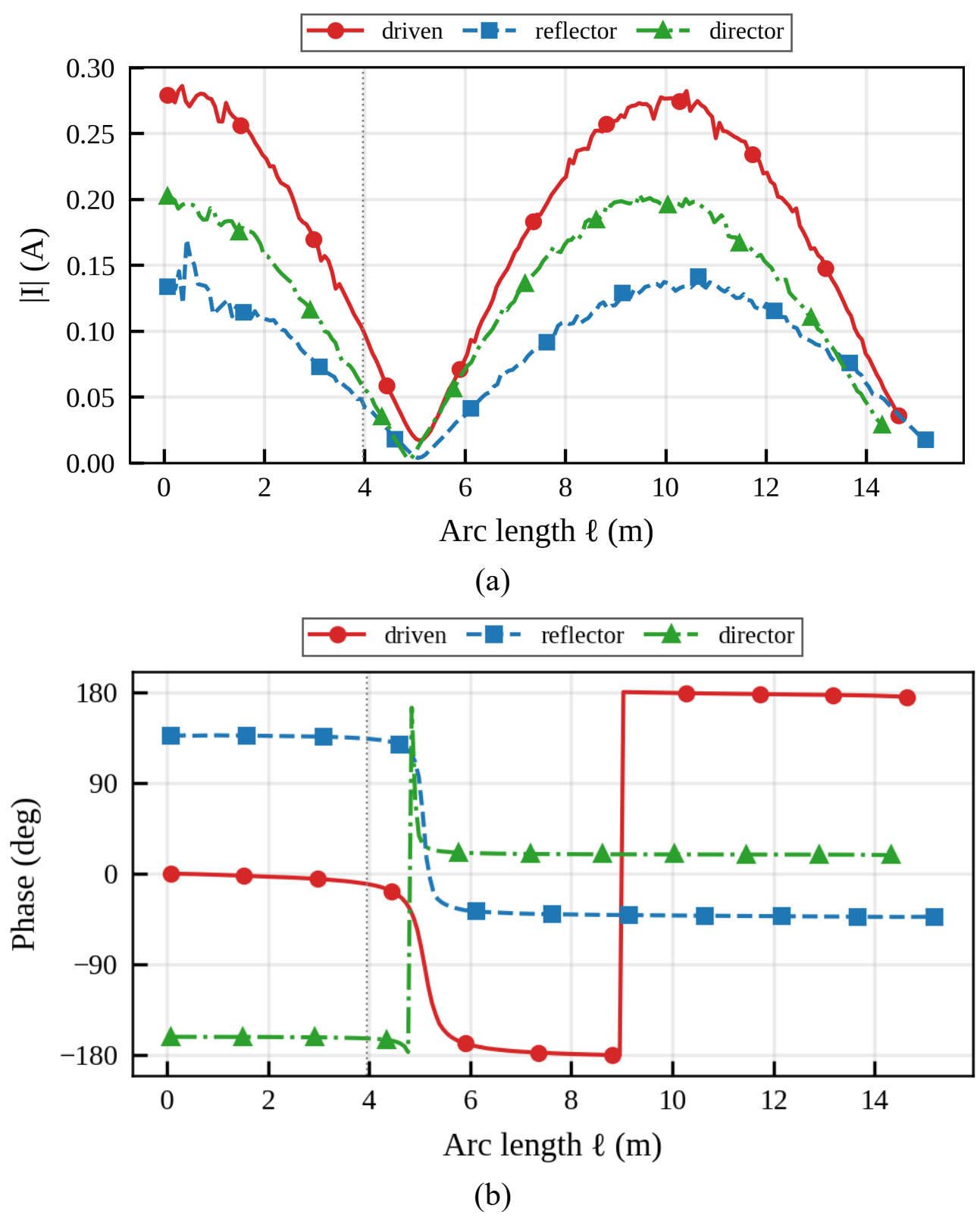

**FIGURE 10.** Extracted currents (60-cm lines) of the BO-optimized 3-element curved array (candidate A): magnitude and phase relative to the driven element's feed current. The dotted line marks the end of the tube.

## VI. VALIDATION

### A. CONVERGENCE OF THE CURRENT EXTRACTION

Fig. 11 shows the current of the single curved element extracted with offsets of 6, 15, 30, and 60 cm. The 6-cm and 15-cm lines underestimate the current, by up to 50% on the arc, because they lie within a few mesh cells of the thin wire, where the field is poorly resolved. Between 30 and 60 cm, the extraction converges: the RMS difference is 2.2% on the tube and 5.1% on the arc, and the phase agrees within 1°. The converged feed current (0.190 A) matches the value derived from the port quantities (0.193 A). For every array, the feed current from (15) agreed with $\sqrt{2P_{\mathrm{acc}}/R_{\mathrm{in}}}$ within 1%–5% (e.g., 0.310 against 0.311 A for the sweep-optimized 3-element array and 0.279 against 0.275 A for candidate A).

The converged current also answers the question of how sinusoidal the current really is. The magnitude agrees with $I_0 \sin k(L-\ell)$ within about 5%–10% along most of the element; the arc's peak current is 0.174 A against 0.184 A predicted (95%); and the main difference is a partly-filled current null near 5 m (0.033 A instead of zero), i.e., a small traveling-wave component caused by radiation and ground loss [29]. The sinusoidal model used in the single-element analysis was thus a good approximation, and the array results use the extracted currents directly.

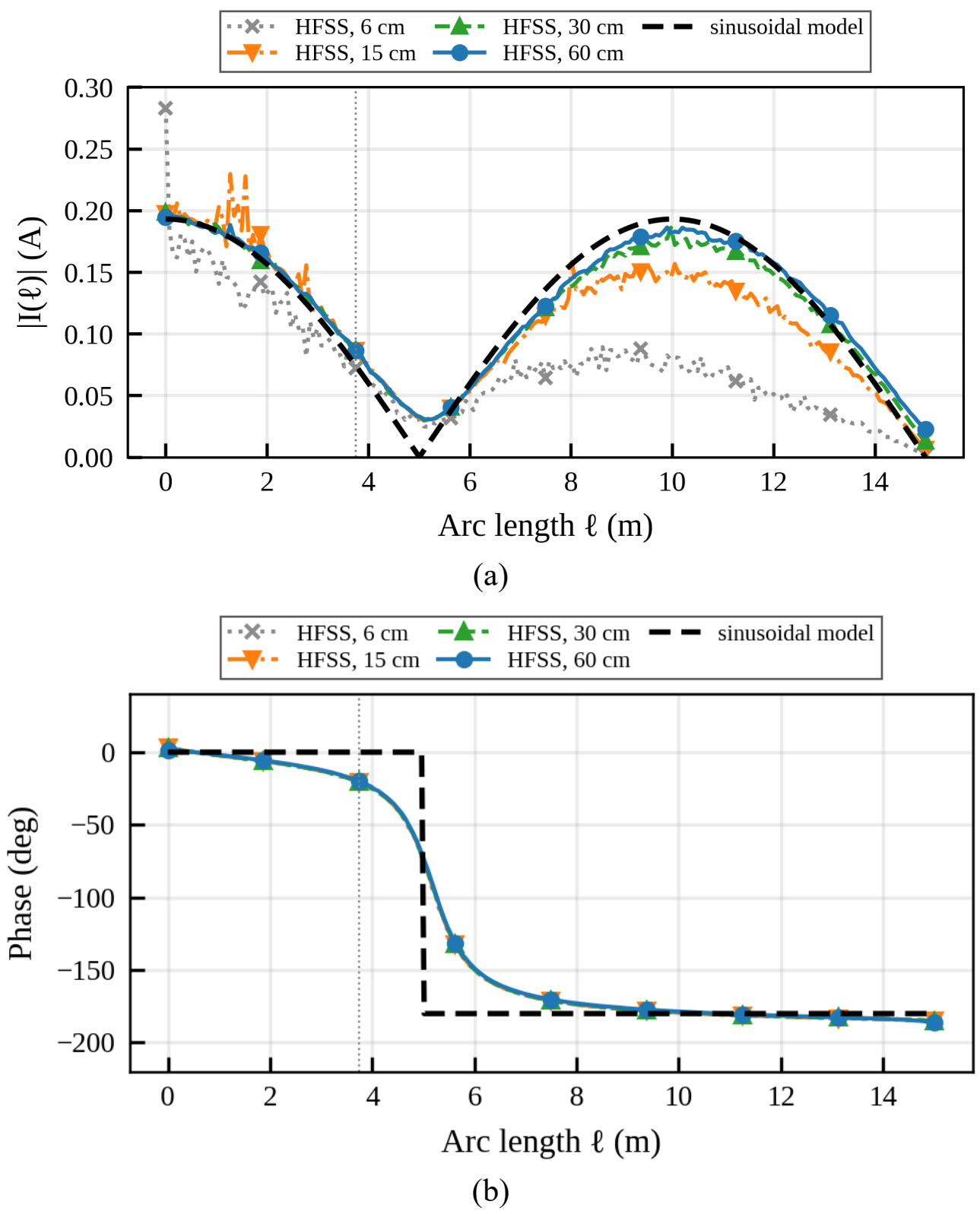

**FIGURE 11.** Current of the single curved element ($L = 15$ m) extracted from HFSS with line offsets of 6–60 cm, compared with the sinusoidal model: (a) magnitude; (b) phase relative to the feed. The dotted line marks the end of the tube.

### B. AGREEMENT WITH EXACT PERFECT-GROUND SOLUTIONS

For four arrays, separate HFSS runs over an infinite PEC ground were compared with the image-theory far field of the currents extracted from the same runs (Fig. 12 and Table 7). For both straight arrays, which radiate TM fields, the model agrees with HFSS within 0.05 dB from $\psi = 0°$ to 30°. For the curved 3-element array, the agreement is within 0.08 dB from 3° to 40°. For the curved 2-element array, it is within 0.16 dB from 10° to 40°; the larger deviation below 5° occurs where its TE field nearly cancels with its image (Proposition 3(ii)), so that small current errors dominate a small residual. The F/B agrees within 0.4–1.2 dB, because the back lobe is itself a small residual after cancellation. Over PEC, TE radiation carries 86% and 90% of the forward-sector power of the curved 2- and 3-element arrays.

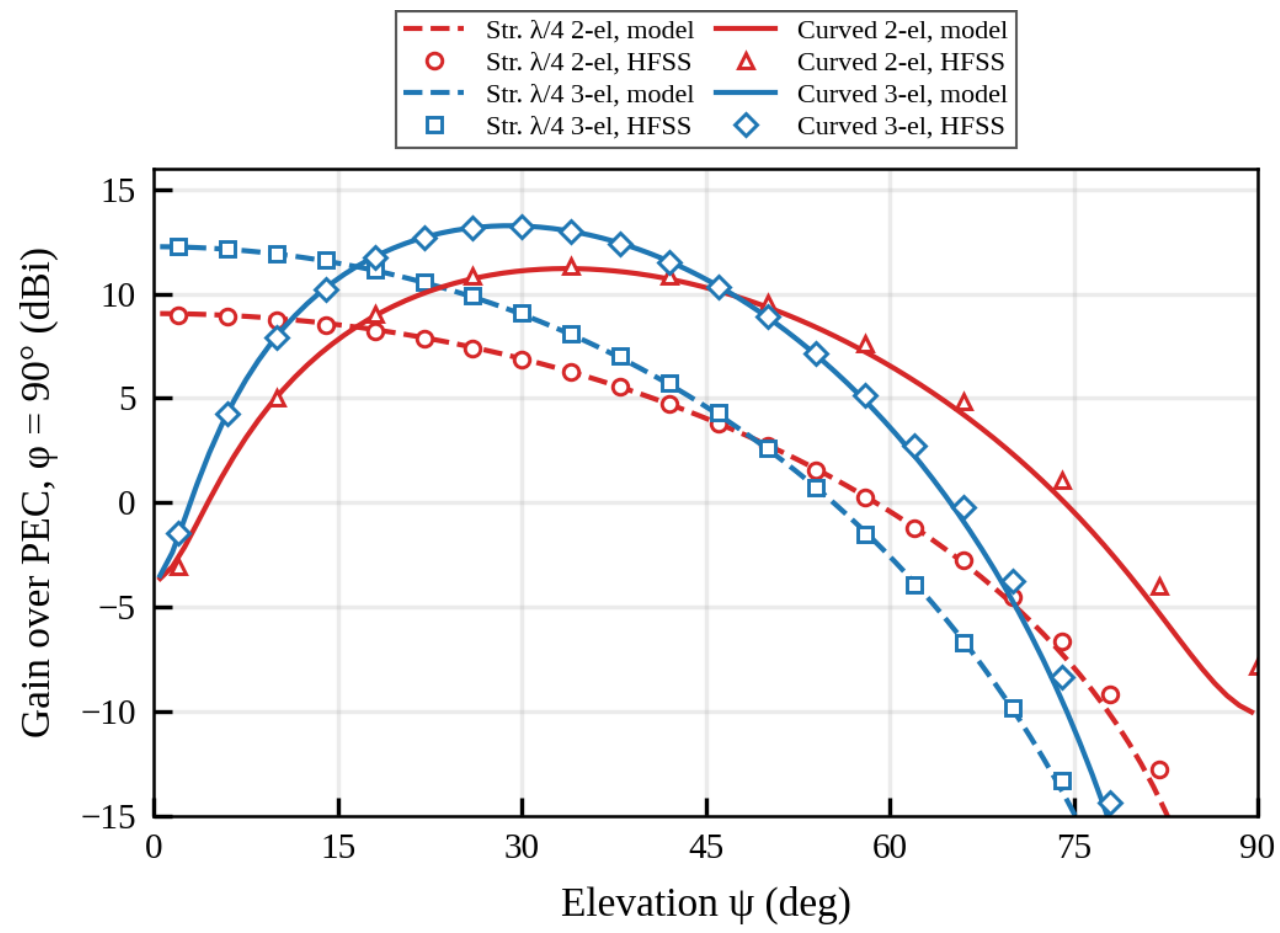

**FIGURE 12.** Validation over an infinite PEC ground: gain at $\phi = 90°$ from the image-theory field of the extracted currents (lines) and HFSS's exact far field (markers) for two straight and two curved arrays.

**TABLE 7.** SUMMARY OF VALIDATION CHECKS

| Check | Design(s) | Result |
|---|---|---|
| Offset convergence, 30 vs. 60 cm | Single curved | RMS 2.2% (tube), 5.1% (arc), phase ≤ 1° |
| Feed current vs. port | All five designs | within 1%–5% |
| PEC far field, ψ = 0°–30° | Straight λ/4 2-el / 3-el | ≤ 0.03 / ≤ 0.05 dB; sector 0.02 / 0.11 dB |
| PEC far field, ψ = 10°–40° | Curved 2-el | ≤ 0.16 dB; sector 0.41 dB* |
| PEC far field, ψ = 3°–40° | Curved 3-el | ≤ 0.08 dB; sector 0.10 dB |
| Power balance | All designs | −0.7% to +3.6% |
| MoM vs. HFSS | Curved 3-el | $R_{\mathrm{in}}$ 13.8 vs. 13.6 Ω; $\overline{G}_f$ 8.87 vs. 8.85 dBi |
| Hemisphere integral of PEC gain | PEC runs | 1.00–1.04 (one faulty run detected) |

*Dominated by near-grazing angles where the TE field cancels with its image.

### C. LIMITATION OF THE NATIVE FULL-WAVE FAR FIELD

The validated post-processing exposes a pitfall of the native far field of full-wave solvers over lossy soil. Table 9 compares the native HFSS forward-sector gain with the realized Fresnel result. For the curved arrays, whose radiation is mostly TE, the two differ by less than 1 dB. For the straight λ/4 arrays, the native far field overestimates the forward-sector gain by 4.3–4.6 dB, because it omits the pseudo-Brewster cancellation that forms beyond the solution box. The same error appears in the curved arrays' native gain at grazing incidence: over PEC, the curved 3-element array gives 0.1 dBi at $\psi = 3°$, but the native lossy-earth solution gives 6.0 dBi, although a real ground reflects horizontal polarization with $\Gamma_h \approx -1$ there. Comparisons of vertical antennas over soil based only on the native far field can therefore favor them by several decibels.

## VII. COMPARISON WITH STRAIGHT ARRAYS

### A. OPTIMIZED STRAIGHT BASELINES

The straight baselines are vertical Yagi–Uda arrays optimized by the same procedure as the curved arrays. For the quarter-wave 2-element array, a 35-run sweep ($k_r = 1.02$–$1.10$, $s_r = 3$–$6$ m) and a driven-length sweep (4.40–4.58 m) gave $k_r = 1.08$ and $s_r = 4.0$ m. Shortening the driven element made it resonant but reduced $R_{\text{in}}$ to 22.6 Ω and the F/B to 7.5 dB, so the 4.58-m element was retained; a straight λ/4 2-element array cannot be simultaneously resonant, self-matched, and high in F/B at this spacing. For the 3-element array, an 18-run director sweep ($k_d = 0.88$–$0.98$, $s_d = 3$–$5$ m; Fig. 8, dashed) gave $k_d = 0.96$ and $s_d = 4$ m with $R_{\text{in}} = 9.4$ Ω, matched by 0.168 µH and 442 pF.

Straight 0.575λ elements, optimal as single radiators over seawater and PEC (Table 1), were also examined. They are non-resonant at 15 MHz ($Z_{\text{in}} \approx 55$–$69 - j215$ Ω, resonances at 10.9 and 18.6 MHz) and, as unloaded parasitic elements, give no reflector or director action: the F/B is about 3 dB for two elements and −0.5 dB for three. Making them work would require base-loading reactances and a large matching network, so they were excluded as array baselines.

### B. EFFECT OF CURVATURE ALONE

To isolate curvature, straight arrays with element lengths and spacings identical to the curved arrays were simulated (Table 8). Curvature alone raises the native forward-sector gain by 3.3 dB (2 elements) and 4.5 dB (3 elements), raises the F/B by up to 6 dB, and lowers the structure by 25%. The straight arrays beam at 46°–52°; over PEC, the straight 3-element array even has a null of −11 dBi at $\psi \approx 20°$, in the middle of the skywave window, because its 0.75λ elements carry opposing current lobes. Curvature removes this null and moves the radiation into the window, which is the pattern effect of Section IV-B; over soil, the TE soil effect adds to it.

**TABLE 8.** Effect of curvature alone (identical element lengths and spacings)

| Array | Ground / model | Curved $\overline{G}_f$ (dBi) | Straight $\overline{G}_f$ (dBi) | Difference (dB) |
|---|---|---|---|---|
| 2-element | Moist, HFSS native | 6.33 | 3.05 | +3.3 |
| 3-element | Moist, HFSS native | 7.65 | 3.14 | +4.5 |
| 3-element | PEC, HFSS exact | 9.07 | 3.83 | +5.2 |
| 3-element | Moist, Fresnel ($\eta = 1$) | 8.85 | 6.62 | +2.2 |

### C. REALIZED GAIN OVER MOIST SOIL

Table 9 and Figs. 13–14 compare the realized gains over moist soil, computed from the extracted currents with (16) and (17). The curved 2-element array reaches 6.75 dBi in the forward sector, 6.4 dB above the optimized straight λ/4 2-element array (0.39 dBi); the BO-optimized curved 3-element array reaches 8.28 dBi, 6.3 dB above the straight λ/4 3-element array (1.97 dBi). In the low-elevation band (3°–10°), the curved arrays exceed the straight ones by 4.2–4.3 dB. The beam peaks are 10.1 and 11.9 dBi at 26°–30° for the curved arrays, against 2.0 and 4.1 dBi at 22°–25° for the straight arrays, and TE radiation carries 81%–85% of the curved arrays' forward-sector power. All arrays are weak at grazing incidence ($\psi < 5°$), as Propositions 1 and 3(ii) predict, so the advantage is concentrated at the elevations of about 8°–30° that dominate skywave coverage.

**TABLE 9.** Realized performance over moist soil (Fresnel post-processing of extracted currents)

| Design | $\overline{G}_f$ 3°–30° (dBi) | $\overline{G}_f$ 3°–10° (dBi) | $G$ at 10°/20°/30° (dBi) | F/B (dB) | TE share | HFSS native $\overline{G}_f$ (dBi) |
|---|---|---|---|---|---|---|
| Single curved | 2.87 | −2.59 | 0.7 / 5.0 / 6.1 | 0 | 77% | 2.91 |
| Curved 2-el | 6.75 | 1.33 | 4.9 / 9.2 / 10.1 | 11.8–14.6 | 81% | 6.33 |
| Curved 3-el (sweep) | 8.51 | 3.44 | 7.7 / 11.7 / 12.1 | 11.5–14.1 | 85% | 7.65 |
| Curved 3-el (BO) | 8.28 | 3.14 | 7.4 / 11.4 / 11.8 | 11.3–13.7 | 85% | 7.50 |
| Straight λ/4 2-el | 0.39 | −2.99 | −0.6 / 1.8 / 1.8 | 9.9–10.9 | 0% | 4.69 |
| Straight λ/4 3-el | 1.97 | −1.04 | 2.0 / 4.1 / 3.6 | 10.8–12.3 | 0% | 6.52 |

Forward sector: 45° ≤ $\phi$ ≤ 135°. Realized gains include $\eta$, mismatch, and L-network loss. F/B at $\psi$ = 3°–30°.

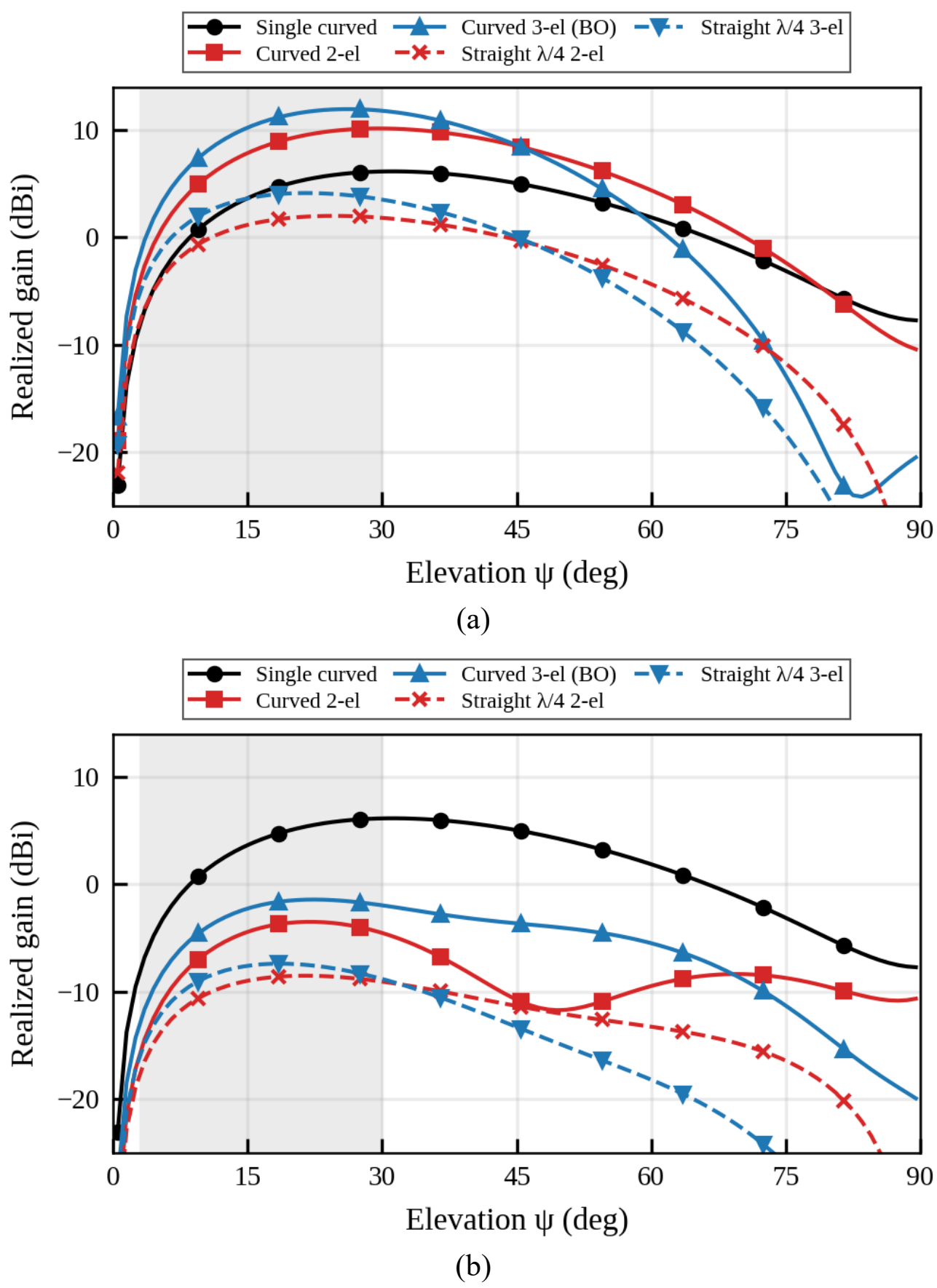

**FIGURE 13.** Realized elevation patterns over moist soil from the extracted currents: (a) forward ($\phi = 90°$) and (b) backward ($\phi = -90°$) cuts. The shaded band marks the 3°–30° sector.

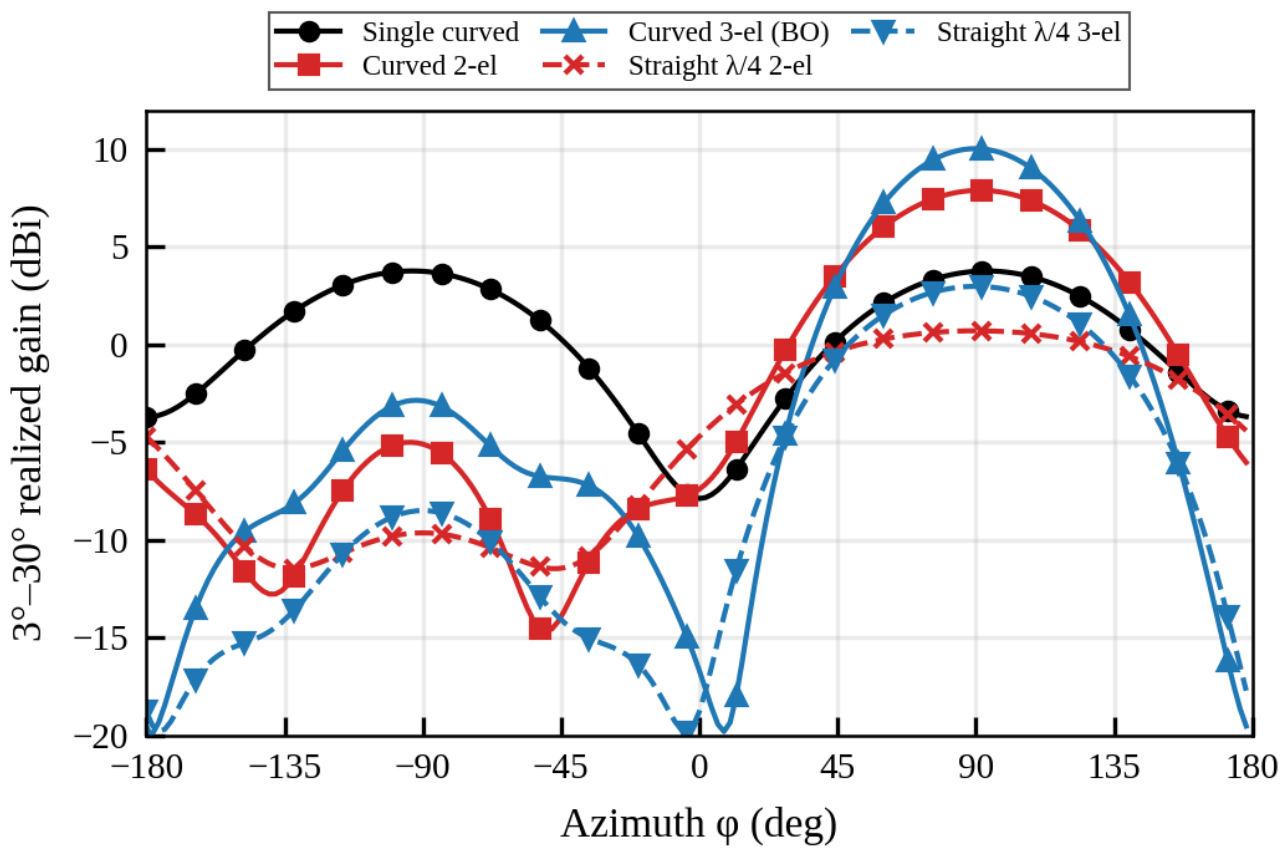

**FIGURE 14.** Realized 3°–30° sector gain versus azimuth over moist soil. The parasitic elements turn the bidirectional TE pattern of the single curved element into a single forward beam.

### *D. SEVEN GROUNDS*

Table 10 and Fig. 15 extend the comparison to seven grounds (pattern-only, $\eta = 1$, so that only the far-field reflection differs). The curved arrays change by only 1.0–1.2 dB from PEC to very dry soil, in agreement with Propositions 2 and 4, whereas the straight λ/4 arrays lose 8.9–9.1 dB. Over PEC and seawater, the straight arrays are better by 0.5–1.2 dB, as Rule 2 predicts; on every land ground, the curved arrays win by 3.4–7.0 dB. Because inland skywave sites experience seasonal changes of soil moisture, the near-independence of the curved arrays from the ground constants is itself a practical advantage.

**TABLE 10.** FORWARD-SECTOR GAIN $\bar{G}_f$ (DBI) VERSUS GROUND (PATTERN ONLY, $\eta = 1$)

| Design | PEC | Sea | Wet | Avg. | Moist | Poor | V. dry |
|---|---|---|---|---|---|---|---|
| Single curved | 3.44 | 3.42 | 3.30 | 3.03 | 3.20 | 2.72 | 2.38 |
| Curved 2-el | 7.20 | 7.18 | 7.09 | 6.83 | 7.01 | 6.54 | 6.18 |
| Curved 3-el (BO) | 8.92 | 8.90 | 8.77 | 8.45 | 8.66 | 8.13 | 7.75 |
| Straight λ/4 2-el | 8.15 | 7.69 | 3.62 | 1.80 | 1.80 | −0.03 | −0.72 |
| Straight λ/4 3-el | 10.09 | 9.61 | 5.43 | 3.56 | 3.57 | 1.68 | 0.96 |

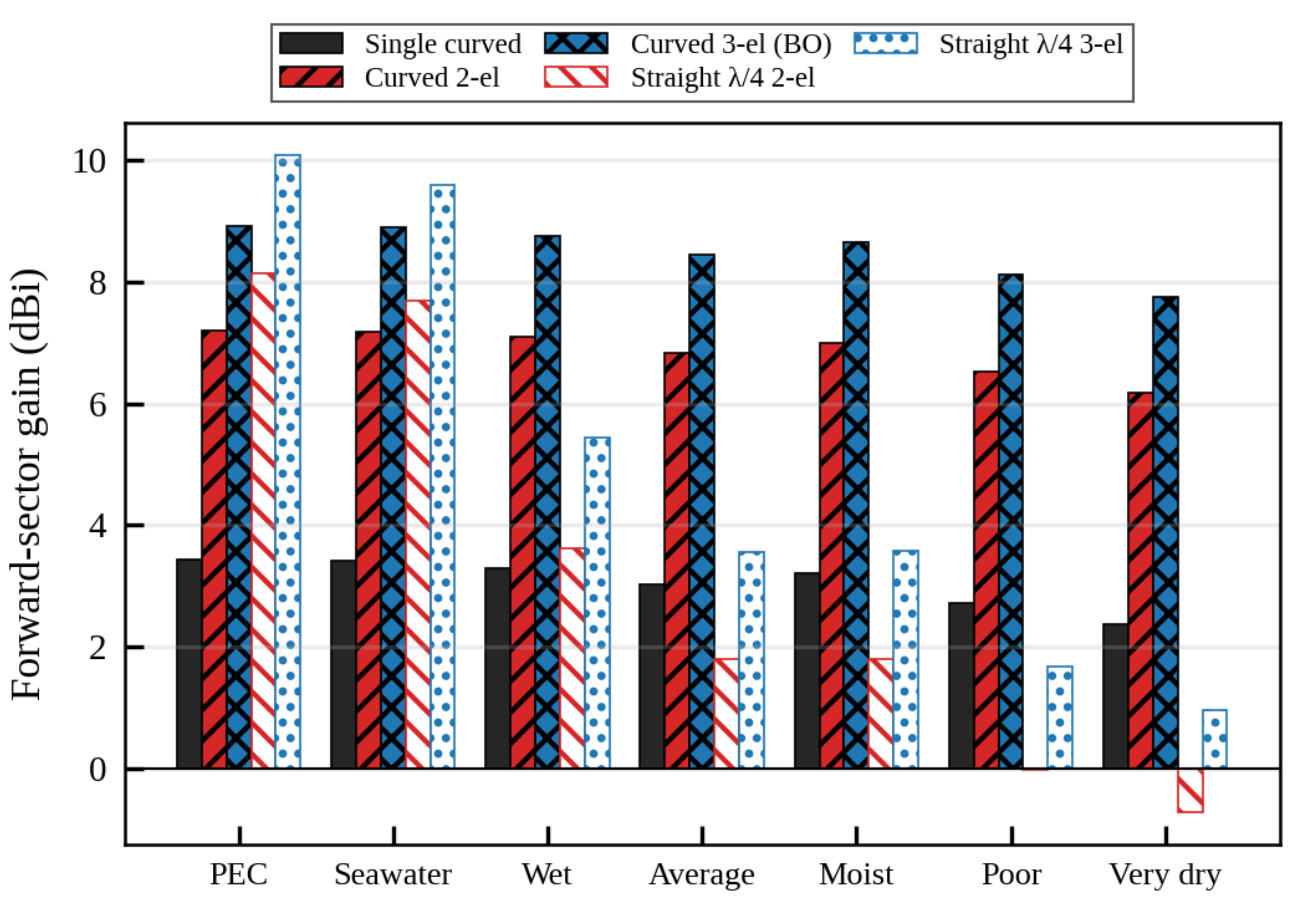

**FIGURE 15.** Forward-sector gain of the five designs over seven grounds (pattern only, $\eta = 1$). Hatched bars: straight λ/4 arrays.

### *E. EFFICIENCY, MATCHING, AND BANDWIDTH*

Fig. 16 and Table 11 summarize the power balances and the impedance properties. The curved designs radiate 92%–95% of the accepted power into the air, whereas the straight λ/4 arrays radiate only 70%–78%: their strong vertical currents near the ground drive more current through the soil, mostly beyond the screen ($P_{\text{far}} \approx 2.7 P_{\text{near}}$). The efficiency difference alone is worth 0.85 dB (2 elements) and 1.2 dB (3 elements). The 2-element arrays are self-matched; the curved 2-element array has the deeper match (−22.4 against −11.4 dB) and the straight one the slightly wider band (8.2% against 6.8%; Fig. 17). The 3-element arrays require L-networks with negligible loss (0.02–0.05 dB). Their matched bandwidths, 1.67% for the curved BO design and 2.05% for the straight array, are typical of tightly coupled 3-element Yagi–Uda arrays and well within the Bode–Fano limit [30], which allows about 4.4% for $Q = 62$ and 6.8% for $Q = 40$ at −10 dB; a double-tuned network could widen them further.

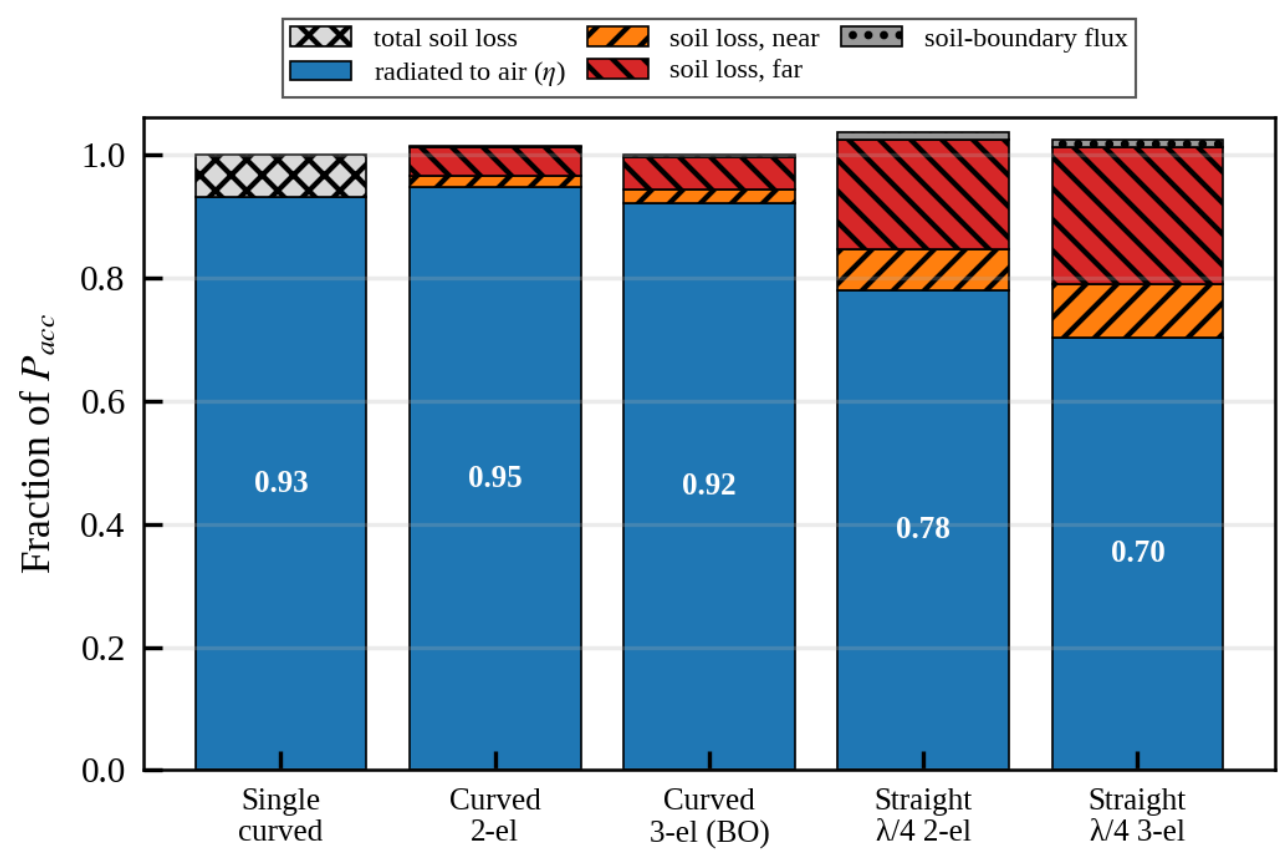

**FIGURE 16.** Power balance over moist soil: fraction of the accepted power radiated into the air ($\eta$) and lost in the near and far soil regions or through the soil boundaries. For the single element only the total soil loss is shown.

**TABLE 11.** IMPEDANCE, MATCHING, BANDWIDTH, AND EFFICIENCY (FINE 25-KHZ SWEEPS)

| Design | $Z_{\text{in}}$ (Ω) | Matching | $\lvert S_{11} \rvert$ (dB) | −10-dB BW (%) | $Q$ | $\eta$ |
|---|---|---|---|---|---|---|
| Single curved | 53.7 + j4.3 | none | −25.3 | 6.0 | – | 0.930 |
| Curved 2-el | 57.4 + j3.4 | none | −22.4 | 6.8 | – | 0.947 |
| Curved 3-el (sweep) | 13.8 − j0.2 | 0.24 µH, 343 pF | −45.3 | 1.06 | 62 | 0.935 |
| Curved 3-el (BO, A) | 21.6 + j7.1 | 0.188 µH, 244 pF | −51.3 | 1.67 | 40 | 0.921 |
| Curved 3-el (BO, B) | 25.0 − j1.5 | 0.281 µH, 213 pF | −49.1 | 2.14 | 31 | 0.934 |
| Curved 3-el (BO, C) | 14.2 − j3.2 | 0.273 µH, 337 pF | −44.4 | 1.08 | 61 | 0.934 |
| Straight λ/4 2-el | 31.2 + j11.3 | none | −11.4 | 8.2 | – | 0.78 |
| Straight λ/4 3-el | 9.4 + j3.7 | 0.168 µH, 442 pF | −45.0 | 2.05 | – | 0.70 |

$\lvert S_{11} \rvert$ with the matching network where used (coil quality factor 150).

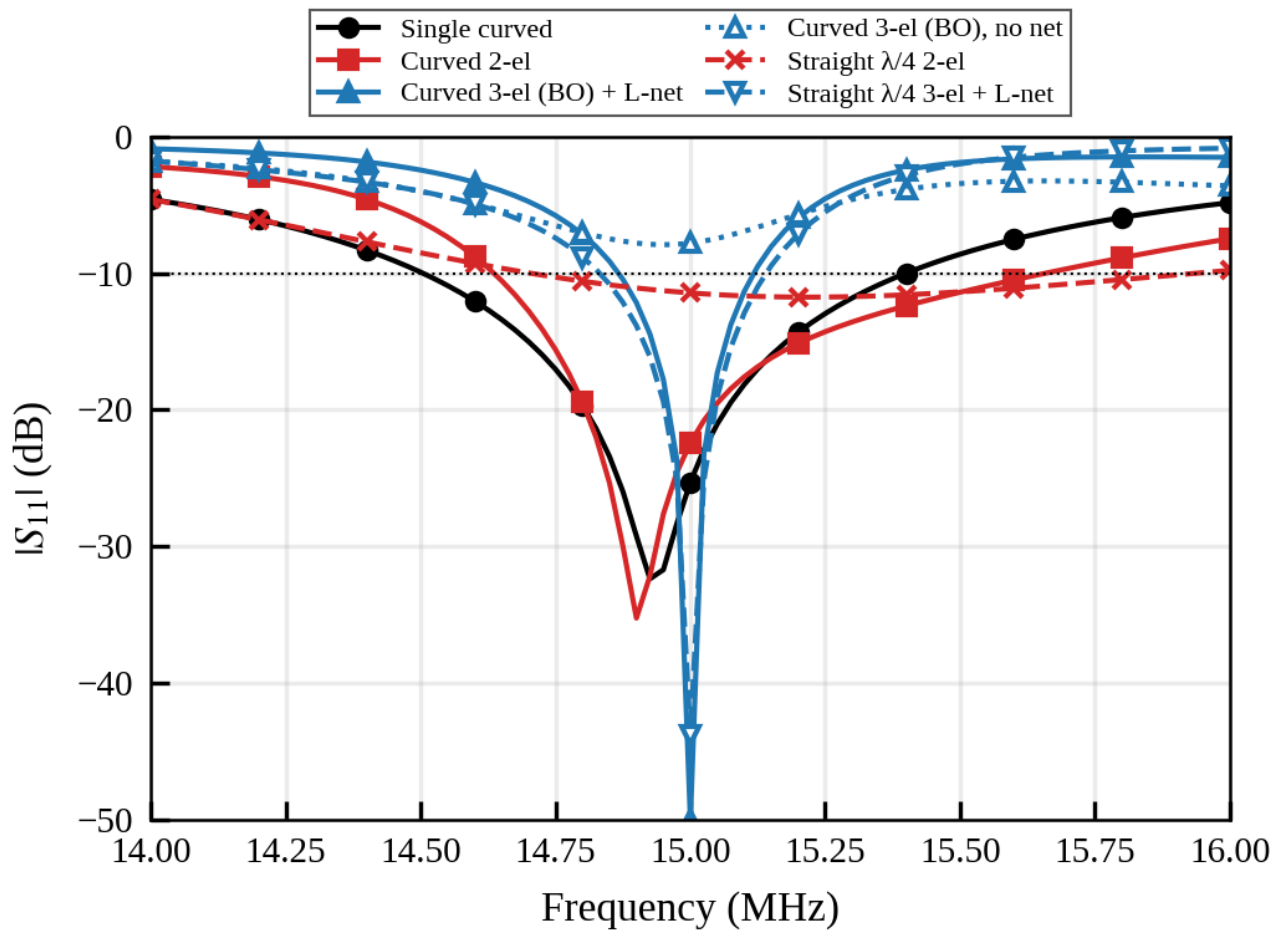

**FIGURE 17.** Fine-sweep $|S_{11}|$ of all designs; the 3-element arrays are shown with their L-networks (coil quality factor 150) and, for the BO design, also without.

## VIII. DISCUSSION

### A. ROBUSTNESS TO CONSTRUCTION ERRORS

Each curved design was perturbed by ±2% in one dimension at a time (length, tube ratio, arc angle, reflector and director factors, and spacings), with the matching network of the 3-element arrays held fixed (Fig. 18 and Table 12). Single-dimension errors change the gain at 10° by at most 0.10, 0.23, 0.30, and 0.18 dB for the single element, the 2-element array, the sweep-optimized 3-element array, and candidate A, respectively, and the worst $|S_{11}|$ remains −20.7, −18.4, −12.3, and −19.3 dB. Uniform ±2% scaling of all elements, which is electrically equivalent to a ±2% frequency shift, is tolerated by the single element and the 2-element array ($|S_{11}| \leq -10.6$ dB), although the latter's F/B falls to about 6 dB at −2%, so the F/B, not the gain or the match, sets the usable band of about 14.8–15.2 MHz. For the 3-element arrays, uniform scaling exceeds the matched bandwidth and requires an adjustable matching network, as is standard for HF Yagi–Uda arrays. Candidate A halves the sensitivity to single-dimension errors and the gain loss under +2% scaling (0.86 against 1.63 dB at 10°), which confirms quantitatively that the lower Q found by BO has practical value.

**TABLE 12.** Robustness to ±2% construction errors

| Design | Max \|ΔG\| single dim. (dB) | Worst $\|S_{11}\|$ single dim. (dB) | ΔG uniform −2% / +2% (dB) | $\|S_{11}\|$ uniform −2% / +2% (dB) |
|---|---|---|---|---|
| Single curved | 0.10 | −20.7 | +0.12 / +0.07 | −17.8 / −10.6 |
| Curved 2-el | 0.23 | −18.4 | −0.52 / +0.28 | −13.0 / −12.8 |
| Curved 3-el (sweep) | 0.30 | −12.3 | −0.44 / −1.63 | −3.5 / −1.7 |
| Curved 3-el (BO) | 0.18 | −19.3 | −0.16 / −0.86 | −5.3 / −3.4 |

ΔG at $\psi = 10°$, $\phi = 90°$ (HFSS native). Fixed L-networks for the 3-element arrays.

### B. PRACTICAL IMPLICATIONS

Because the gain stems from polarization conversion, the curved arrays suit skywave paths, where magnetoionic splitting into ordinary and extraordinary modes makes the total radiated power the relevant measure [31], but not surface-wave radars, which require TM radiation [4]. For a monostatic radar using the same antennas on transmit and receive, the forward-sector advantages of 6.3–6.4 dB translate into about 12.6–12.8 dB of two-way gain over the optimized straight λ/4 arrays on moist soil. The curved arrays are taller (11 m against 4.9 m), which is the main cost; they are, however, 25% lower than straight arrays with the same wire lengths, need no matching network in the 2-element version, and use identical masts and arc formers for all elements. The 3-element arrays need an L-network, preferably with a variable capacitor to re-center the match after construction or when changing channels. In phased arrays of curved Yagi–Uda elements, the TE beams of the elements should face the scan sector.

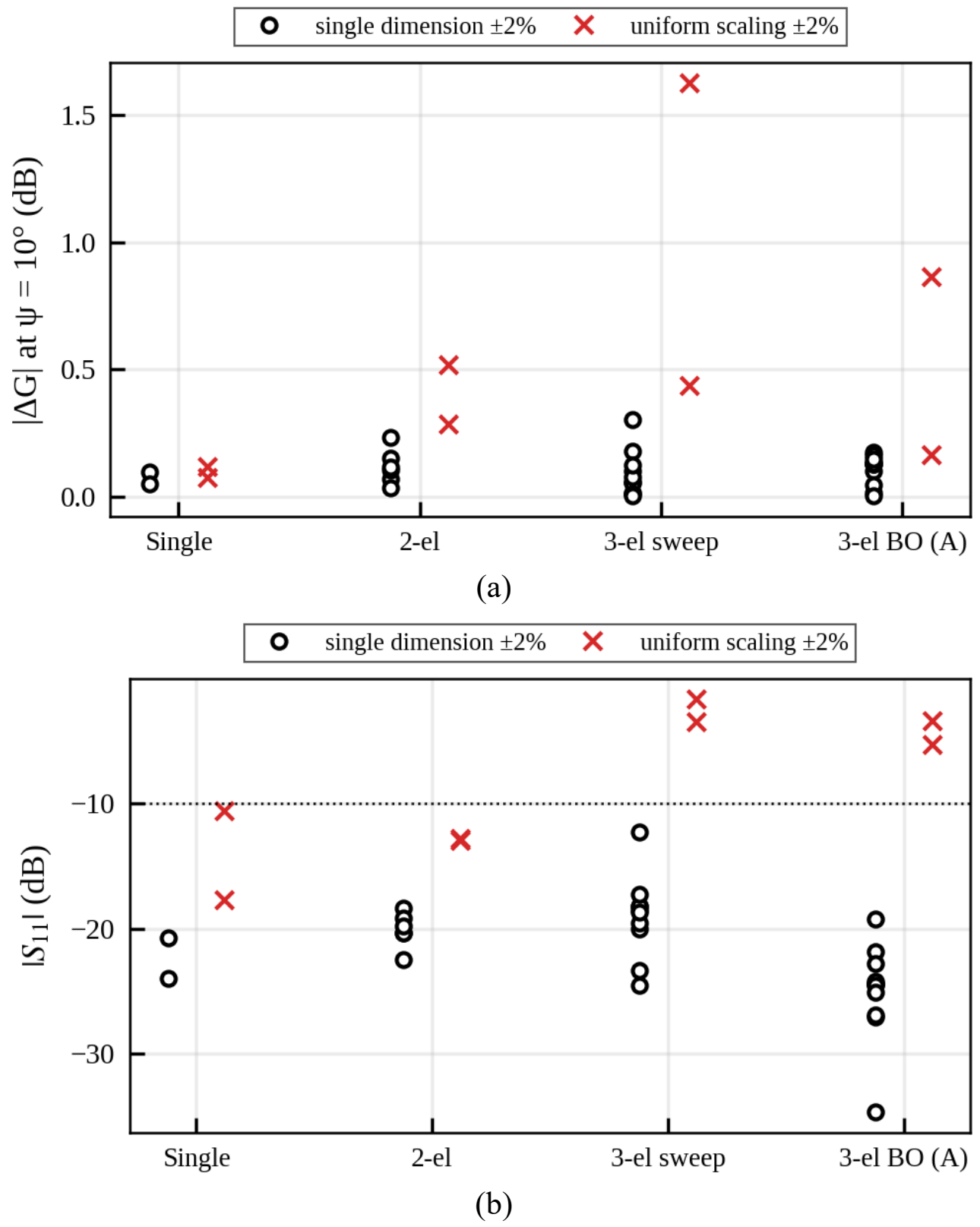

**FIGURE 18.** Robustness to ±2% errors: (a) gain change at $\psi = 10°$ and (b) $|S_{11}|$ (with the fixed matching network for the 3-element arrays) for single-dimension errors (dots) and uniform scaling (crosses).

### C. LIMITATIONS AND PLANNED MEASUREMENTS

The present results are computational. They rest on full-wave site simulations, a validated current extraction, and the reflection-coefficient approximation, which excludes the Norton surface wave and assumes a homogeneous half-space; layered soils can be treated by replacing (3) and (4) with layered reflection coefficients [13]. A measurement campaign is being prepared on scaled prototypes. At a 1:20 scale (300 MHz), the curved elements are about 0.73 m long and 0.55 m tall, so that the 2- and 3-element arrays fit on a 0.82-m ground plane in an anechoic chamber, where $S_{11}$, the elevation and azimuth patterns, and the F/B over a perfect ground can be measured and compared with the PEC results of Section VI-

B. Because the soil conductivity cannot be scaled exactly with frequency, the lossy-ground behavior will be verified indirectly by comparing measured currents or patterns over the ground plane with the validated model, and, where feasible, by full-scale tests at an inland HF site.

## IX. CONCLUSION

This paper has shown, through analysis and validated full-wave design, that curving the elements of HF monopoles and Yagi–Uda arrays redirects radiation from a soil-limited TM route into a TE route that is only weakly affected by soil. Closed-form bounds quantify both routes, and a separability result explains why arrays whose elements share a mast and an arc circle inherit the single element's soil insensitivity. We developed a hybrid methodology combining Bayesian optimization on analytical and MoM models with full-wave site simulations. The current extraction converged to an RMS difference of at most 5.1%, the power-balance errors ranged from −0.7% to +3.6%, and Fresnel post-processing agreed with exact perfect-ground references within 0.16 dB pointwise for four arrays (with a 0.41-dB sector-mean deviation for the curved 2-element array). The resulting self-matched 2-element and BO-optimized 3-element curved Yagi–Uda arrays achieve realized forward-sector gains of 6.8 and 8.3 dBi over moist soil, 6.4 and 6.3 dB above straight $\lambda/4$ arrays optimized in the same way, with 92%–95% efficiency. Their gain varies by only about 1 dB from a perfect ground to very dry soil, whereas that of the straight arrays falls by 9 dB. At equal wire length, curvature alone contributes 3.3–4.5 dB of gain and permits a 25% lower structure. Bayesian optimization on the validated MoM surrogate revealed a gain–bandwidth Pareto front that HFSS confirmed at three points within 0.1 dB: its high-gain end exceeds the best parametric-sweep design by 0.2 dB at nearly equal bandwidth, and the selected balanced design widens the matched bandwidth by 58% and halves the sensitivity to construction errors at a cost of 0.23 dB. The results also show that the native far field of full-wave solvers can overestimate the low-angle gain of vertical arrays over soil by more than 4 dB. Scaled measurements and multi-element phased arrays of curved Yagi–Uda elements are the next steps.